\documentclass[aps,letterpaper,showpacs,showkeys,amsmath,amssymb,tightenlines]{revtex4}
\usepackage{natbib}
\usepackage{latexsym}
\usepackage{graphicx}
\usepackage{amsmath}
\usepackage{amssymb}
\usepackage{bm}
\usepackage{float}
\usepackage{color}
\bibpunct{(}{)}{,}{a}{}{,}

\def\gtrless{\raise2.5pt\hbox{$>$}\llap{\lower2.5pt\hbox{$<$}}}
\def\gtrapprox{\raise2.5pt\hbox{$>$}\llap{\lower2.5pt\hbox{$\approx$}}}
\newcommand{\gl}[1]{Eq. (\ref{#1})}
\newcommand{\gls}[2]{Eqs. (\ref{#1},\ref{#2})}
\newcommand{\glto}[2]{Eqs. (\ref{#1}) to (\ref{#2})}
\newcommand{\bsq}[1]{\begin{subequations}\label{#1}}
\newcommand{\esq}{\end{subequations}}
\newcommand{\beq}[1]{\begin{equation}\label{#1}}
\newcommand{\eeq}{\end{equation}}
\newcommand{\beqa}[1]{\begin{eqnarray}\label{#1}}
\newcommand{\eeqa}{\end{eqnarray}}
\newcommand{\disc}[1]{\textcolor{magenta}{}}
\newcommand{\com}[1]{\textcolor{red}{}}
\newcommand{\add}[1]{#1}
\newcommand{\rem}[1]{}
\newcommand{\mike}[1]{#1}
\newcommand{\mikecom}[1]{}
\newcommand{\rev}[1]{#1}
\newcommand{\remrev}[1]{}

\newcommand{\gd}{\dot{\gamma}}
\newcommand{\smop}{\Omega}
\newcommand{\smopb}{\Omega^{\dagger}}
\newcommand{\smoe}{\Omega_e^{\dagger}}
\newcommand{\smoa}[1]{\Omega_a^{\dagger}(#1)}
\newcommand{\smor}[1]{\Omega_r^{\dagger}(#1)}
\newcommand{\smoq}[1]{\Omega_Q^{\dagger}(#1)}
\newcommand{\smos}[1]{\Omega_\Sigma^{\dagger}(#1)}
\newcommand{\smoi}[1]{\Omega_i^{\dagger}(#1)}
\newcommand{\dod}{\delta\Omega^{\dagger}}
\newcommand{\dodb}{\overline{\delta\Omega^{\dagger}}}
\newcommand{\kap}{\boldsymbol{\kappa}}
\newcommand{\rb}{{\bf r}}
\newcommand{\bp}{\boldsymbol{\partial}}
\newcommand{\qb}{{\bf q}}
\newcommand{\kb}{{\bf k}}
\newcommand{\pb}{{\bf p}}

\renewcommand{\rho}{\varrho}
\renewcommand{\epsilon}{\varepsilon}

\begin{document}
\bibliographystyle{plainnat}

\title{A mode coupling theory for Brownian particles in homogeneous steady shear flow}

\pacs{82.70.Dd, 83.60.Df, 83.50.Ax, 64.70.P-, 83.60.Fg}
\keywords{Colloids, Nonlinear rheology, Glass transition, Steady
shear, Non-equilibrium}

\author{M. Fuchs}\affiliation{Fachbereich Physik, Universit\"at Konstanz,
 78457 Konstanz, Germany
}

\author{M. E. Cates}\affiliation{SUPA, School of Physics and Astronomy, The University of Edinburgh,
Edinburgh EH9 3JZ,  UK}

\date{\today}

\begin{abstract}
A microscopic approach is presented for calculating general
properties of interacting Brownian particles under steady shearing.
We start from exact expressions for shear-dependent steady-state
averages, such as correlation and structure functions, in the form
of generalized Green-Kubo relations. To these we apply
approximations inspired by the mode coupling theory (MCT) for the
quiescent system, accessing steady-state properties by integration
through the transient dynamics after startup of steady shear. Exact
equations of motion, with memory effects, for the required transient
density correlation functions are derived next; these can also be
approximated within an MCT-like approach. This results in closed
equations for the nonequilibrium stationary state of sheared dense
colloidal dispersions, with the equilibrium structure factor of the
unsheared system as the only input. In three dimensions, these
equations currently require further approximation prior to numerical
solution. However, some universal aspects can be analyzed exactly,
including the discontinuous onset of a yield stress at the ideal
glass transition predicted by MCT. Using these methods we
additionally discuss the distorted microstructure of a sheared
hard-sphere colloid near the glass transition, and consider how this
relates to the shear stress. Time-dependent fluctuations around the
stationary state are then approximated, and compared to data from
experiment and simulation; the correlators for yielding glassy
states obey a `time-shear-superposition' principle. The work
presented here fully develops an approach first outlined previously
(M. Fuchs and M. E. Cates, Phys. Rev. Lett. {\bf 89}, 248304,
(2002)), while incorporating a significant technical change from
that work in the choice of mode coupling approximation used, whose
advantages are discussed.

\end{abstract}
\pacs{82.70.Dd, 83.60.Df, 83.50.Ax, 64.70.Pf, 83.10.-y}

\maketitle

\section{Introduction}
\label{intro}

 Colloidal dispersions represent one of the
simplest classes of materials for which the interplay between
viscoelasticity and externally controlled flow can be investigated.
Quiescent dispersions consisting of colloidal, slightly polydisperse
(near-)hard spheres exhibit all the hallmarks of a glass transition
as the volume fraction is increased. At densities above this
transition, Brownian motion of the colloids is ineffective in
relaxing structural correlations on observable timescales: the
system remains amorphous, like the fluid phase, but becomes
nonergodic. The colloidal glass transition has been studied in
detail by dynamic light scattering measurements
\citep{Pus:87,Meg:91,Meg:93,Meg:94,Heb:97,Bec:99,Bar:02,Eck:03},
confocal microscopy \citep{Wee:00} and both linear
\citep{Mas:95,Zac:06} and nonlinear rheology
\citep{Pet:99,Pet:02,Pet:02b,Pha:06,Pha:08,Bes:07,Cra:06,Cra:08}.
The nonlinear rheology of colloidal glasses appears, at least
macroscopically, to be characterized by the appearance of dynamic
yield-stress behavior in which (a) any finite steady shear rate
restores ergodicity, and (b) a finite limiting stress is attained on
slowly reducing the shear rate towards zero.

Although there is some consensus about the mechanisms (normally
described in terms of cage-formation) that control the colloidal
glass transition, it is not clear {\em a priori} whether these also
control the nonlinear rheology of dense suspensions. For example,
many nonlinear effects have been attributed to ordering or layering
of the particles \citep{Lau:92} and/or cluster formation
\citep{Bes:07,Ben:96,Gan:06}. Either could be important, especially
under conditions where hydrodynamic interactions dominate. This
dominance seems relatively unlikely at the infinitesimal shear rates
seen in colloidal glasses just beyond yield (at least, not if the
flow remains homogeneous, which we assume here). Several theoretical
studies have suggested a connection between steady-state nonlinear
rheology and the glass transition, but apart from our own work, few
of these theories explicitly address colloids. For instance, the
mean field approach to spin glasses was generalized to systems with
broken detailed balance in order to model `flow curves' of glasses
\citep{Ber:00,Ber:02}; but although the microscopic model is clear,
the relation to actual shear flow proceeds only by analogy. The soft
glassy rheology model, which describes mechanical deformations and
ageing \citep{Sol:97,Sol:98,Fie:00}, explicitly addresses shear but
contains almost no structural information on the material under
study.

Conversely, in a colloidal context, rheological models have grown up
largely without reference to the glass transition. For example,
\citet{Bra:93} worked out a scaling description of the rheology \rev{of colloids} 
based on the concept that the structural relaxation arrests at
random close packing (RCP). This contains important insights, but if
colloidal arrest is actually at the glass transition (which for hard
sphere colloids is observed at volume fractions of about 58\%, well
below RCP at 64\%), it gives in the end a \mike{potentially}
misleading picture. This is because almost everything one could
measure diverges at RCP (osmotic pressure, local lubrication
resistance, shear modulus...) whereas at a glass transition there is
only one divergent quantity: the structural relaxation time. This
discrepancy therefore cannot be fixed by a simple rescaling of the
volume fraction to make the RCP point coincide with the observed
arrest density.

In contrast, the mode coupling theory (MCT) of the glass transition
gives broadly the right sort of divergence in relaxation times (as
probed by scattering experiments), without unwanted singularities in
pressure or short-time dynamics, but does so at a density that is
too low by several percent. The results are traditionally shifted to
recover the correct arrest point, whereafter MCT provides a
near-quantitative explanation of numerous dynamic measurements at
the glass transition in quiescent colloidal dispersions. Two
important effects are however neglected; one is aging
\citep{Pur:06}, the other, residual activated decay processes at
ultra-long times that may cause any glass to flow ultimately
\citep{Goe:91,Goe:92,Goe:99}. Neglect of activated rearrangements
defines the so-called `ideal' glass transition; it is this ideal
limit that standard MCT addresses and that we wish to extend here to
the case of steady shear. 
\rev{The ultimate flow regime caused by activated hopping would presumably convert the yield stress that we find below for the ideal case into a regime of extremely high but finite viscosity (caused by a very large finite relaxation time). So long as colloid experiments do not access these ultimate time scales, the yield-stress scenario that we develop here should be a good description.}

An appealing aspect of the standard MCT is
that its sole input is the static structure factor or equal-time
density correlator $S_q$ in the equilibrium state (see below).
\rev{This structure factor is used to create an approximate expression for the thermodynamic forces that arise when particles adopt a given Fourier-space density pattern. A separate assumption of our approach is that, at a time in the distant past prior to onset of shearing, the system was governed by the Boltzmann distribution. This assumption is of course correct for ergodic fluids; however, in the glass it represents only one of many possible protocols
governing the preparation of the system. Because the system starts in equilibrium, slow aging processes (which in practice lead towards that state from a nonequilibrium initial state) are precluded from our description. In the context of the present work, which addresses steady-state shear only, this is almost certainly not important since, as we shall see, steady shear restores ergodicity. Thus we expect no dependence on the choice of initial state. However, to check this with certainty, calculations would have to be performed with a different initial ensemble. MCT-like techniques to address aging by this route (without shear) have been developed, but encounter notable technical difficulties \cite{latz}, and we do not pursue them here.}

Alongside the present work, several MCT-inspired approximations have
been used to describe the nonlinear rheology of colloidal
dispersions. 
\citet{Ind:95} described
self-diffusion at low densities, where they suggested a
non-selfconsistent, perturbative solution.
\citet{Miy:02} made this approach self-consistent, extended it to
collective density fluctuations in dense fluids close to, but always
below, the glass transition. (\citet{Miy:04,Miy:06} also presented a
field theoretic derivation, evaluated their equations
quantitatively, and tested the results by computer simulations and
experiments.)  These approaches investigated the time-dependent
fluctuations around the stationary state under shear,\mike{ and in
principle required as input the distorted structure factor (albeit
then approximating this by the undistorted one).}\disc{[IS THIS
REALLY TRUE? IF SO, HOW WAS IT CALCULATED?]. Yes, in philosophy, but
not in actual calculation. They quoted Rhonis' result for it, and
then used (claiming it to be a technical approximation) that the
distorted is close to the quiescent one.} In spirit, they followed
\mike{closely the original MCT without shear.} However, these
authors' wariness of addressing the rheology of the glass phase
itself was warranted; their theory invokes a fluctuation-dissipation
theorem that cannot be relied upon in the glass. Finally, in the
interesting recent approach of \citet{Kob:05,Sal:08}, entropic
barrier hopping prevents glass formation in an `extended MCT'
framework, and applied stresses modify the barrier heights.

In assessing this progress, it is worth recalling some of the prior
history of MCT. This has, broadly speaking, been used to deduce the
dynamics from equilibrium structural information in three
situations. \citet{Kawasaki}  considered phase transitions, where
critical fluctuations lead to self-similar scaling laws in the
structure functions. \citet{Goe:91,Goe:92}  developed MCT for
glasses, where the equilibrium structure varies smoothly but a
bifurcation arises in the equations of motion. Long time tails and
back flow phenomena could also be described by rather similar
equations, where, however, in contrast to glasses, only hydrodynamic
long wavelength fluctuations are important \citep{Kawasaki1973b}. A
central tenet for work in these three areas has been that the
equilibrium structural information used in the mode coupling
equations should be under control and well understood. For the
nonlinear steady-state rheology of a dense fluid close to arrest
into an amorphous solid, the {\em equilibrium} structure structure
factor $S_q$ is, by definition, unchanged from the quiescent case
and thus under control. But the same does not hold for the
steady-state equal-time correlator in the flowing system or
`distorted structure factor'. At a small fixed flow rate, the latter
quantity, just like the shear stress (to which it is closely
related) can be qualitatively different just outside and just within
the glass phase. Indeed, if the glass transition is accompanied by
the abrupt onset of a finite yield stress, both the stress and the
correlator should have a discontinuity, in the limit of
infinitesimal flows, at the glass transition. We shall find that
this is indeed a prediction of our approach. (The result is not
obvious; for instance the SGR model \citep{Sol:97} has a yield
stress that rises smoothly from zero on entering the glass, giving
no such discontinuity.) Thus, for any theory of the nonlinear
rheology of the glass transition, a central issue is the handling of
the stationary structural correlations. We therefore devote part of
this paper (Section \ref{disc}) to discussing these in detail.

\mike{Our} integration through transient (ITT) approach, first
suggested \rev{by} \citet{Fuchs2002c}, takes a somewhat different route to
the challenge of extending MCT to address the rheology of dense
fluids close to the glass transition. The method proceeds by
studying the build up of structural correlations under the combined
influence of flow and Brownian motion, after switch-on of steady
shearing. The initial state, even in the glass, is taken to be the
equilibrium one, governed by the Boltzmann distribution. (This is
also how standard MCT proceeds -- for quiescent systems, the ideal
glass transition is then defined by the loss of ergodicity within an
initial Boltzmann state.) ITT differs from all the aforementioned
approaches to MCT under shear by focussing on the {\em transient}
density correlators, which follow from equations of motion
containing the equilibrium structure factor $S_q$ as input. The
equilibrium $S_q$ is determined by the interaction pair potential
among the particles, and is assumed to vary smoothly with
thermodynamic control parameters. Within quiescent-state MCT, $S_q$
is in fact used as proxy for the pair potential, to calculate
thermodynamic forces on the particles. ITT follows this avenue
directly, rather than making any parallel assumption about the
nonequilibrium equal-time correlator (the distorted structure
factor). This is physically advisable since the relation between
structure factor and particle interactions only holds in the absence
of shear: the structure factor is distorted because the system is
driven away from Boltzmann equilibrium, not because it adopts the
Boltzmann distribution of some distorted pair potential.

The distorted structure factor is then \mike{not} an input but an
output of the ITT approach, just like the stress (to which it is
closely related). Importantly, we will find -- as anticipated above
-- that the distorted structure factor is nonanalytic in density at
the glass transition for all finite shear rates, and nonanalytic in
shear rate throughout the glass. Within ITT, time-dependent
fluctuations around the stationary state can be \mike{computed},
but, in contrast with the other approaches, do not play a central
role. Specifically their time integrals do not give the transport
coefficients, such as viscosity, that characterize the steady state.
In fact, stationary correlation functions and dynamic
susceptibilities {\em can} be connected via extended forms of the
fluctuation-dissipation theorem, but the familiar and useful
versions that apply to linear response around equilibrium states are
violated. (These violations have recently been studied by
\citet{Kru:08} using approximations beyond those outlined here.)

Having surveyed the relation to other approaches in the literature,
we now summarize the relationship between \mike{the work of this
article} and our own previous publications on this topic. The first
of these was a short paper \rev{\citep{Fuchs2002c}} outlining in
preliminary form: (i) the route via Green-Kubo formulae to exact
equations that form the basis of the ITT approach; (ii) the use of
projection methods on these to obtain MCT-like approximations to
them in the form of closed equations for transient correlators;
(iii) the resulting bifurcation structure; (iv) development of
semi-schematic and fully schematic MCT models inspired by the closed
equations (which \mike{otherwise} remain intractable in three
dimensions); and (v) numerical results from these schematic models.
The semi-schematic (ISHSM) and fully schematic
($F_{12}^{(\dot\gamma)}$) models were subsequently elaborated \rev{by}
\citet{Fuchs2003} with additional results, variants and experimental
comparisons appearing in several subsequent papers:
\citet{Fuc:03b,Fuc:04,Hen:05,Cra:06,Cra:08,Haj:08}. In the present
work we invoke an ISHSM model in Section \ref{disc} (when addressing
quantitatively the physics of yielding and of the distorted structure
factor) which differs marginally from the previously published
version, as detailed in Appendix C. But otherwise, we do not
rehearse any material relating to items (iv,v) above. \mike{(}For
completeness in Section \ref{disc} we do however briefly restate the
results of \citep{Fuchs2002c} and \citep{Fuchs2003} on the
bifurcation structure, item (iii).\mike{)}

The major goal of this paper is to give a full exposition of both
the ITT formalism prior to its approximation by MCT, and the MCT
approximations proper: that is, items (i) and (ii) in the above
list. However, in developing our MCT-based approximations, we make a
significant technical change to the ones used originally by
\citet{Fuchs2002c} \rev{and} \cite{Cates2004a,Fuchs2005a}. Within the present scheme,
which involves a different definition of the transient correlator
$\Phi_{\bf q}(t)$ from the one in \citep{Fuchs2002c}, the initial
decay rate $\Gamma_{\bf q}(t)$ in the correlator memory equation
(\gl{e21} below) is guaranteed positive, not only in the quiescent
state, but also under shear. It seems desirable to retain this
property, since the MCT approach was originally developed under
conditions where $\Gamma_{q}>0$, and both its physical adequacy and
its numerical stability are unproven for other cases. Moreover, the
resulting formulae are generally simpler and more elegant, at least
if we neglect certain additional terms which now arise in the memory
equation, whose form we discuss. Both schemes reduce to standard MCT
in the absence of flow, so we are free to make this revised choice
of approximation. An intriguing consequence of doing so is that it
increases the mathematical similarity between our mode-coupling
vertex under shear and that proposed
by  \citet{Miy:02} \rev{and} \cite{Miy:04,Miy:06}. Given the very different precepts
of the two approaches (as detailed above) this does not however
imply any deeper equivalence of their theory and ours.

Our altered choice of MCT closure brings about some mathematical and
notational changes which have only a minor effect on the basic
structure of the theory,  but are pervasive and sometimes subtle.
For this reason, although \rev{\citet{Fuchs2005a}} already presented
the exact stages of the ITT formalism (item (i)) using the
previously defined correlator, we re-work much of this material here
with the new definition. This has the advantage of making the
current paper more self-contained, although we still refer to
\rev{\citet{Fuchs2005a}} for some important technical results that do not
depend on the choice of definition made.

The work presented here on steady states underpins two recent short
papers in which we announce extensions of our MCT approach to deal
with nonsteady shear \citep{Bra:07} and to general unsteady flows
\citep{Bra:08}. (The first of these uses the original correlator
definition by \rev{\citet{Fuchs2002c}}; the second uses the definition
adopted here.) These rest even more heavily on the  ITT approach
than does the present work, and are made possible because the
integration through transients need not assume constant flow rate,
nor need the integration continue to infinite times. Both
simplifications are \mike{however} retained in this paper which
concerns only the steady state, long after shear startup. The
notational overhead of presenting the ITT method for time-dependent
flows is considerable, and we have ourselves found the full theory
to be much easier to understand once the steady-state version is
mastered. Therefore we restrict attention to steady shear in this
work.

In common with most MCT-based approaches, we entirely neglect the
hydrodynamic interactions that stem from the presence of an
incompressible solvent surrounding \mike{our Brownian} colloidal particles. A
partial justification for this is the hope that, close to a glass
transition, \mike{the main effect of hydrodynamics} is to renormalize the timescale of
local diffusive transport (as characterized by the bare diffusion
constant, $D_0$). Since at the transition the structural relaxation
time is a divergent multiple of this local time (as a result of the
increasing difficulty, and ultimate failure, in escaping from local
cages) any smooth density dependence of $D_0$ caused by hydrodynamic
interactions is probably unimportant. (Note that in the quiescent
state, hydrodynamic interactions also cause the Brownian motion of
individual particles to become correlated, but this does not change
the argument.) Under flow, solvent incompressibility also requires
locally large (but zero-mean) deviations in velocity from that
imposed macroscopically. Such deviations could quantitatively
influence all our results but, for low enough shear rates, need not
harm the qualitative picture that emerges.
\rev{This describes the yielding and shear-melting of the amorphous solid or glass. It does not describe any form of shear-thickening -- a phenomenon which certainly can arise in many dense colloidal suspensions, albeit primarily at high bare Peclet number rather than the small ones considered here. Thus our theory should be viewed throughout as a low-shear rate approximation; however its validity is not limited to the range of linear response. (Indeed, this range shrinks to zero at the glass transition and remains there throughout the glass itself.) While hydrodynamic interactions are clearly implicated in some forms of shear thickening, in others its role is less clear. An example of the latter is the formation under shear of stable arrested granules which then continue to exist on cessation of flow \citep{catesgran}. Elsewhere we have discussed modifications of MCT that can capture these non-hydrodynamic forms of shear thickening, but these remain rather ad-hoc and we do not pursue them here \citep{Hol:03,Hol:05}.}

More dangerous is the possibility of macroscopic inhomogeneities in
flow rate, for instance to form coexisting layers of glassy and
fluid material, \mike{at equal stress but} possibly at slightly
different densities \citep{Bes:07,Var:04,Ben:96,Gan:06,Bal:08}.
However, such phenomena also arise in other fluids such as wormlike
micelles \citep{Cat:06}, and in these cases modelling proceeds by
first assuming a uniform flow and then analysing the resulting
continuum rheology for potential flow instabilities. This justifies
the approach taken here which addresses homogeneous shearing only.
The resulting flow curves (explored \rev{by} \citet{Fuchs2003} \rev{and} \citet{Haj:08})
generally remain monotonic unless deliberately altered (e.g., to
account phenomenologically for shear thickening
\citep{Hol:03,Hol:05}). This monotonicity rules out the most obvious
source of shear banding instabilities, but does not preclude those
involving either coupling to concentration gradients, or
intrinsically unsteady flow \citep{Cat:06}. Both avenues merit
further study, particularly in view of the recent experimental
observations by \citet{Bal:08}, which do suggest macroscopic flow
inhomogeneity under steady shearing in dense colloidal suspensions.

The rest of this paper is organized as follows. Section \ref{model}
details our microscopic starting point and Section \ref{ITT} the
exact manipulations that lead to the ITT methodology. Section
\ref{tradens} addresses the transient density correlators and the
derivation via MCT of their approximate equations of motion. Section
\ref{disc} gives a discussion that includes our new results for the
correlator decay and the distorted structure factor. Section
\ref{conc} gives our conclusions; the Appendices contain some
technical details omitted from the main text.

\section{Microscopic starting point}
\label{model}

The system considered consists of $N$ spherical particles (diameter
$d$) dispersed in a volume $V$ of solvent with imposed flow profile
${\bf v}(\rb) = \kap\cdot  \rb$,  where for simple shear with
velocity along the $x$-axis and its gradient along the $y$-axis, the
shear rate tensor is $\kap =\gd\  \hat{\bf x} \hat{\bf y}$ (that is,
$\kappa_{\alpha\beta}=\gd \delta_{\alpha x}\delta_{\beta y}$).  The
effect of the shear rate $\gd$ on the particle dynamics is measured
by the Peclet number, Pe$_0=\gd d^2/D_0$, formed with
the (bare) diffusion coefficient $D_0$ of a single particle \citep{russel}.
Dimensionless quantities are obtained by using $d$ as unit of
length, $d^2/D_0$ as unit of time, and $k_BT$ as unit of energy,
whereupon Pe$_0 = \gd$. The evolution of the distribution function
$\Psi(\Gamma)$
 of the particle positions,
$\rb_i$, $i=1,\ldots, N$ (abbreviated into $\Gamma=\{{\bf r}_i\}$),
under internal forces ${\bf F}_i = - \bp_i U(\Gamma)$ (with the
total interaction potential $U$) and shearing, but neglecting
hydrodynamic interactions, is given by the Smoluchowski equation
\citep{russel,dhont}: \beqa{b1}
\partial_t \Psi(\Gamma,t) & = & \smop(\Gamma) \; \Psi(\Gamma,t)\; ,
 \nonumber\\
\smop & = & \Omega_{e} + \delta \Omega = \sum_i \bp_i \cdot \left(
\bp_i - {\bf F}_i - \kap \cdot \rb_i \right)  \; ; \eeqa here
$\Omega_{e}= \sum_i \bp_i \cdot \left( \bp_i - {\bf F}_i \right)  $
abbreviates the Smoluchowski Operator (SO) without shear. In the
following, operators act on everything to the right, if not marked
differently by bracketing. Neglect of hydrodynamic interactions
implies that we are considering a set of Brownian particles, each of
which has mobility $\mu = D_0/k_BT$; a particle at ${\bf r}$ feels a
`flow force' ${\bf v}({\bf r})/\mu$, which for isolated particles
exactly replicates the effect of advection, in addition to the
interaction force ${\bf F}_i$ from other particles.

There exist two special time--independent distribution functions,
the equilibrium one, $\Psi_e$, and the stationary one, $\Psi_s$, which satisfy respectively:
\beq{b2}
 \Omega_e \; \Psi_e =
 0 \qquad , \qquad \smop \; \Psi_s = 0 \; .
\eeq
The equilibrium one is determined from the total internal
interaction energy $U$ via
the Boltzmann weight, $\Psi_e(\Gamma) \propto e^{- U(\Gamma)}$,
as seen from the useful relation:
$\bp_i \, \Psi_e =  {\bf F}_i\, \Psi_e$.
The stationary distribution function $\Psi_{s}$ is, however,
unknown. Equilibrium averages with $\Psi_e$ will be abbreviated by $ \langle
\ldots \rangle = \int  \Psi_e(\Gamma) \ldots d\Gamma$, while
$\Psi_s$ determines steady state averages, denoted by
 $\langle \ldots\rangle^{(\gd)}= \int  \Psi_s(\Gamma) \ldots d\Gamma$.

The adjoint of the SO can be found from partial integrations (using
the incompressibility condition, ${\rm Trace}\{\kap\}=0$) as:
\beq{b3} \smopb = \sum_i ( \bp_i + {\bf F}_i + \rb_i\cdot \kap^T)
\cdot \bp_i = \smoe + \dod\; , \eeq where boundary contributions
will be neglected throughout. We will generally be concerned with
the thermodynamic limit of $V\to\infty$ at fixed particle density
$n=N/V$, but where boundary conditions are required, we assume these
to be periodic.\mikecom{This choice hopefully dispels any
misconception that a flux-free steady state might be found under
shear or any other steady flow: a particle flux is unavoidably
present. Accordingly there is no nonequilibrium free energy, that
is, no $\tilde U$ for which $\Psi_s(\Gamma) \propto e^{+ \tilde
U}$\add{, so that averages can easily be obtained from derivatives
of $\tilde U$}.}\disc{I changed the formula, which we need to
discuss. I believe there is a $\hat U$, so that $\Psi_s\propto
e^{-\hat U(\Gamma)}$; just because $\Psi_s>0$, and we can take the
logarithm. But there is no nonequilibrium free energy $\tilde U$, so
that $\Psi_s(\Gamma) = e^{\tilde U - \hat U(\Gamma)}$, and averages
etc can be obtained from ${\tilde U}$ by differentiation. does this
make sense?}\mikecom{This was about some papers Joe found that
pretended one could replace an extensional flow by an effective
potential. But I am not so concerned about it since those papers are
obviously wrong (they were cited in his first draft of our PRL, but
we took it out for final version). I think the last sentence above
SHOULD be removed since the sign choice is possibly confusing and it
doesn't illuminate the present work. The sentence before it COULD
also be removed, ending the paragraph at 'periodic'.}

\rev{The operator $\smop$ acts to the right on a probability density to give the divergence of the resulting probability flux, which for $\smop = \smop_e$ vanishes in the Boltzmann steady state. Accordingly the latter is a right eigenfunction with eigenvalue zero. The adjoint operator $\smopb$ has that same interpretation when acting to the left, but acting to the right it represents the flux of a gradient. For $\smop = \smop_e$ there is again a right eigenfunction with eigenvalue zero, but this time the eigenfunction is a constant. In practice the transition from a representation involving $\smop$ to one involving $\smopb$ is achieved by partial integration, and is similar to going from a Schr\"odinger to a Heisenberg representation in quantum mechanics. This allows one to work with operators that act on the functions of coordinates whose averages are being taken, rather than acting on the probability densities themselves. By this route one obtains equations relating averages of different quantities taken within standard (Boltzmann or steady-state) distributions. Such equations then invite closure approximations. Closure is instead much harder among equations in which operators alter the distribution functions, even if the two formulations contain equivalent information when handled exactly.} 

\rev{Note further that the manipulations carried out in this Section make no assumption about whether the system is glassy -- although that does of course inform the choice of approximation made subsequently. Rather, we are concerned here with exploiting the invariance properties of steady states. Time translation invariance restricts the form of static and dynamic correlators to those discussed below, whereas translational invariance can also be exploited, as usual, by transforming from particle coordinates to Fourier components of the density. The same holds for any spatially fluctuating quantity, including the microscopic stress tensor, whose zero-wavevector component is the macroscopic stress.}

The shear-dependent operator $\dod$ will be shown to capture the
affine distortion of density fluctuations under shear. It is given
by \beq{b4} \dod= \sum_i \bp_i \cdot \kap \cdot \rb_i = \sum_i \rb_i
\cdot \kap^T \cdot \bp_i \eeq where  ${\rm Trace}\{\kap\}=0$ was
again used. Without applied shear the SO $\smopb_e$ is a Hermitian
operator with respect to equilibrium averaging \citep{dhont}
\beq{b5} \langle g\; \smopb_e\; f^* \; \rangle =\langle f^* \;
\smopb_e\;  g \rangle = - \sum_{i}\; \langle  \frac{\partial
f^*}{\partial \rb_i} \cdot
 \frac{\partial g}{\partial \rb_i} \rangle\; ,
\eeq and (as seen from specialising to $f=g$) possesses a negative
semi-definite spectrum. Here and in the following\mike{,} symbols
like $f$, $g$, etc., denote arbitrary functions of the full set of
particle positions: $f=f(\Gamma)$. With shear, however, $\smopb$
cannot be brought into a Hermitian form \citep{Risken}. This follows
from the presence of particle fluxes in the steady state. Such
fluxes violate time reversal symmetry and there is no detailed
balance principle; the latter asserts the cancellation of all
microscopic fluxes for a quiescent system in steady state.
\rev{Recently, it has been shown for the present situation, that this entails anomalous fluctuation dissipation ratios at long times \citep{Kru:08}.}

The action of $\smop$ on the equilibrium distribution function
$\smop \Psi_e=\delta \smop \Psi_e$ can be written in terms of the
stress tensor that arises from the interparticle forces: \beq{b6}
\delta \smop\,
 \Psi_e =
- \sum_i \bp_i \cdot \kap \cdot \rb_i \; \Psi_e = - \sum_i {\bf F}_i
\cdot \kap \cdot \rb_i  \; \Psi_e =
  {\rm Trace}\!\left\{\kap \cdot {\boldsymbol \sigma} \right\}\;  \Psi_e = \gd \ \sigma_{xy}\; \Psi_e ,
\eeq where the specific form of $\kap$ for \mike{simple shear flows}
was used in the last equality only. In \gl{b6},
$\sigma_{\alpha\beta}$ is the zero--wavevector  limit of the
potential part of the stress tensor, defined as: \beq{b7}
\sigma_{\alpha\beta} = - \sum_i \;  F^\alpha_i r^\beta_i \; . \eeq
It is crucial to our approach that the shear stress enters the
calculations in two distinct ways: one is the obvious one (as a
rheological quantity worthy of study), and the other is as a
generator of the transformation between equilibrium and
nonequilibrium averages. The latter role stems from \gl{b6} and
comes to the fore in subsequent developments.

As detailed in the Introduction, we address only homogeneous (and
amorphous) systems so that, by assumption, the stationary
distribution function  $\Psi_s$ remains translationally invariant
even when it becomes anisotropic as a result of shearing.
\citet{Fuchs2005a} showed that this assumption is compatible with
use of the SO in \gl{b1}, even though the latter appears to break
translational invariance. (The proof rests on arguments equivalent
to those leading to \gl{b10} below.) At finite shear rate, we shall
consider below wavevector-dependent fluctuations around the steady
state, $\delta f_\qb = f_\qb - \langle f_\qb \rangle^{(\gd)}$, and
obtain for these quantities not only steady-state averages, but also
time-dependent correlation $C_{fg;\qb}(t)$ and time-independent
structure functions $S_{fg;\qb}$ (definitions to follow). 
Translational invariance brings appreciable simplifications for all
such averages involving time-independent functions of the
time-varying coordinates $\Gamma$: \beq{b8} f_{\bf q}(\Gamma,t) =
e^{\smopb \, t} \; \sum_i\; X^f_i(\Gamma) \; e^{i\qb \cdot {\bf
r}_i}\; .  \eeq Important examples include $X^\rho_i=1$ which
describes density variations ($f_{\bf q}(\Gamma,t) =\rho_{\bf
q}(t)$), and \beq{} X^{\sigma_{\alpha\beta}}_i =\delta_{\alpha\beta}
+ \frac{1}{2} \sum_{j\ne i} (r^\alpha_i-r^\alpha_j) \frac{du(|{\bf
r}_i-{\bf r}_j|)}{dr^\beta_i}\; ,\eeq \disc{[WHY PRIME ON SUM?], to
state $j\ne i$}from which we obtain the full wavevector-dependent
stress tensor ($f_{\bf q}(\Gamma,t) =\sigma_{\alpha\beta}({\bf
q},\{{\bf r}_i(t)\})$) for particles at positions $\{{\bf r}_i(t)\}$
interacting via a pair potential $u(r)$. \rev{While the wavevector
dependence of density fluctuations is quite familiar, the one of
stress fluctuations is e.g.~discussed by \citet{Balucani1994}.}
\rev{As stated previously, the purpose of shifting from real space to Fourier space is to exploit translational invariance which means that equal time correlators form a diagonal matrix in $\bf q$
space but not in real space. For unequal times, there is a similarly important but subtler simplification: correlators can only connect wavevectors that are advected into one another by the intervening flow.
We next explore in turn these consequences of translational invariance for equal and unequal times.}

Translational invariance in an infinite sheared system dictates that
averages involving such quantities are independent of identical
shifts of all particle positions: $\rb_i' = \rb_i + {\bf a}$ for all
$i$, which we denote as $\Gamma \to \Gamma'$. Under such a shift the
SO becomes \beq{b9} \smopb(\Gamma) = \smopb(\Gamma') - {\bf P }
\cdot \kap \cdot {\bf a} \; , \quad \mbox{ with }\, {\bf P} = \sum_i
\bp_i \; . \eeq Thus, for any fluctuation of a variable which
depends on particle separations only,
 viz.  $X^f_i(\Gamma) = X^f_i(\Gamma')$, we have ${\bf P} X^f_i(\Gamma)=0$. From this it follows that \beq{b10} f_\qb(\Gamma,t) = e^{- i \left( \qb +
\qb \cdot \kap \; t \right)\cdot {\bf a} } \; f_\qb(\Gamma',t) \; ,
\eeq whose proof uses the fact that $\smop$  and ${\bf P}\cdot \kap
\cdot {\bf a}$ are commuting operators \citep{Fuchs2005a}. (This in
turn holds because the shear rate tensor satisfies $\kap\cdot \kap =
0$, and because the sum of all internal forces vanishes due to
Newton's third law.) As the integral over phase space must agree
whether integration variables $\Gamma$  or $\Gamma'$ are chosen, it
follows from \gl{b10} that steady-state averages can be nonvanishing
for zero wavevector only: \beq{b12} \frac 1V\; \langle f_{\qb}(t)
\rangle^{(\gd)} = f(\gd)\; \delta_{\bf q , 0} \; .\eeq Note that the
volume $V$ is taken to be finite at first, with periodic boundary
conditions, in order to work with a discrete set of wavevectors and
with Kronecker $\delta$'s. Finally, the thermodynamic limit is
taken. (For consistency, this procedure requires all physical
correlations to be short ranged.) Examples of stationary averages
are the average density $n=N/V$ (which is independent of
$\dot\gamma$) and the macroscopic shear stress $\sigma(\gd) =
\langle \sigma_{xy}\rangle^{(\gd)}/V$.

By similar arguments, wavevector-dependent, anisotropic,
steady-state equal-time correlators, built from pairs of
fluctuations $\delta f_\qb,\delta g_{\qb'}$, are diagonal in
$\qb,\qb'$ indices. Accordingly we define, as one would in a system
at rest, \beq{b13} S_{fg; {\bf q}}(\gd) = \frac 1N \; \langle
\delta f^*_\qb \; \delta g_\qb  \rangle^{(\gd)}\;,\eeq where we
adopt the convention that, where no explicit time arguments are
given for $f$ or $g$, the two times are equal. These `structure
functions' are independent of time in steady state; the familiar
equal-time (distorted) structure factor, built with density
fluctuations, shall be denoted $S_{\bf q}(\gd)= \frac 1N \; \langle
\delta\rho^*_\qb \; \delta\rho_\qb \rangle^{(\gd)}$.

\begin{figure}[t]
\centering
\includegraphics[ width=0.4\columnwidth]{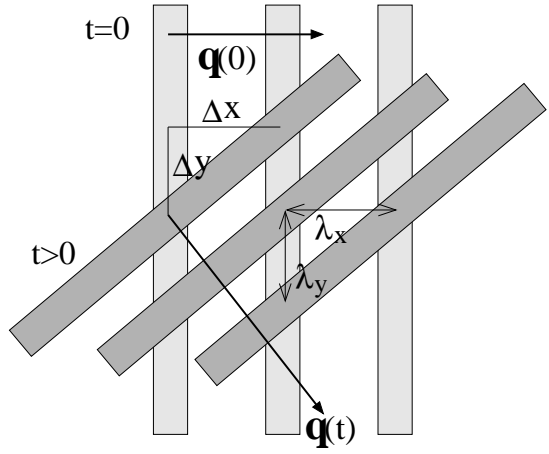}
\caption{Shear advection of a fluctuation with initial wavevector in \mike{the}
$x$-direction, $\qb(t\!\!=\!\!0)=q\, (1,0,0)^T$, and advected
wavevector at later time  $\qb(t\!\!>\!\!0)= q\,  (1,-\gd t,0)^T$.
While $\lambda_x$ is the wavelength in \mike{the} $x$-direction at $t=0$, at
later time $t$, the corresponding wavelength $\lambda_y$ in \mike{the}
(negative) $y$-direction obeys: $\lambda_x/\lambda_y=\Delta x /
\Delta y= \gd t$. At all times, $\qb(t)$ is perpendicular to the
planes of constant fluctuation amplitude. Note that the magnitude
$q(t)=q\sqrt{1+(\gd t)^2}$ increases with time. Brownian motion,
neglected in this sketch, would smear out the
fluctuation.\label{Fig0}}
\end{figure}

The extension of such quantities to unequal times requires
explicit account of the time-dependence of the wavevector of a
fluctuation, as indicated in \gl{b10}. By translational invariance, the only nonzero averages connect a wavevector $\qb$ at (arbitrary) time $t'$ with its advected counterpart $\qb(t)$ at later time $t'+t$ (Fig.\ref{Fig0}).
A correlation function characterizing this, chosen to closely resemble the corresponding equilibrium quantity, is: \beq{b14} C_{fg ; {\bf q}}(t,\gd) = \frac 1N \, \langle
\delta f^*_{\qb} \;  e^{\smopb\, t }\,  \delta g_{\qb(t)}
\rangle^{(\gd)} \; . \eeq
Again by convention the suppressed time arguments ($t'$, say) of $f(\Gamma)$ and $g(\Gamma)$ are arbitrary but equal. The exponential factor is a time evolution operator, discussed further in Section \ref{ITT}, which defers the evaluation of $g$, defined via \gl{b8}, until $t'+t$.
(This time evolution operator does not, of course, act on the $\bf{q}(t)$ label itself.)

In \gl{b14} the advected wavevector is defined as \beq{b15} {\bf
q}(t) = {\bf q} - \qb \cdot \kap\, t = {\bf q} - \gd t\; q_x
\hat{\bf y}\; .\eeq Its magnitude shall be denoted as
$q(t)\equiv\sqrt{q_x^2+(q_y-\gd t\,q_x)^2+q_z^2}$, and its square as
$q^2(t)\equiv(q(t))^2$. This describes, as in Fig.\ref{Fig0}, the
advection of a density fluctuation through a time interval $t$. Note
that our previous formulation \citep{Fuchs2002c} instead utilized
the backwards-advected wavevector ($\qb + \qb \cdot\kap\, t$).
Switching to forward advection eases the interpretation and is more
suited to the  revised mode-coupling approximations that we adopt in
this paper\rev{; see the discussion in Appendix B. (As shown there, this change is not equivalent to merely changing the sign of $\dot\gamma$.)}

For the special case of density fluctuations, for which $f_{\bf
q}(\Gamma,t) = g_{\bf q}(\Gamma,t) = \rho_{\bf q}(t)$, the
abbreviation $C_{\bf q}(t,\gd) = \frac 1N \, \langle
\delta\rho^*_{\qb} \;  e^{\smopb\, t }\,  \delta \rho_{\qb(t)}
\rangle^{(\gd)}$ shall be used below. This defines the intermediate
scattering function of the system under shear. Clearly, the complex
conjugate relation $C_{\bf q}^*(t,\gd)=C_{-\qb}(t,\gd)$ holds in
this case. Because of the inversion symmetry of the SO, which,
within our approach, is inherited by $\Psi_e$ and $\Psi_s$ by
assumption,  $C_{\bf q}^*(t,\gd)=C_{\qb}(t,\gd)$ holds also. This
shows that the stationary intermediate scattering function under
shear is real and symmetric in $\qb$.

The steady state averages and correlators defined above all carry
explicitly the shear rate $\gd$ as an argument. In what follows,
wherever this argument is not given explicitly for steady state (as
opposed to transient) quantities, the shear rate is taken to be zero
so that these refer to the equilibrium state, which we continue to
assume to be homogeneous, and also assume to be isotropic. Thus
$S_{fg;\qb}=\langle \delta f^*_\qb \; \delta g_\qb  \rangle/N$ is
the equilibrium $f$-$g$ structure function; $S_q=\langle \delta
\rho^*_\qb \; \delta \rho_\qb  \rangle/N$ is the equilibrium
structure factor for the density; and $C_{ q}(t)$ is the equilibrium
intermediate scattering function.

\rev{The concept of ``wavevector advection'', introduced above, is crucial to subsequent developments, and merits careful explanation. At one level it is merely a way of subtracting out a trivial effect of flow on density correlations -- as would be present even for a system with no interaction between particles and no Brownian motion (Fig.\ref{Fig0}). Because of wavevector advection, if one were to construct a light scattering experiment for such a system under shear, then to observe the time correlation of the density (a non-decaying correlation in this example) the detector would have to be moved in a specific manner so as to allow for the fact that a particular Fourier-space fluctuation or `speckle' is not stationary but has a deterministic motion in reciprocal space resulting from the shear. (Note that this is strictly a thought experiment; light scattering actually measures an intensity autocorrelator from which the density correlator follows, in unsheared systems, via the Siegert relation \cite{Pus:91}.) Interpreted this way, wavevector advection is a book-keeping device for removing this motion. However, below we shall also use the same term to describe the {\em physical consequences} of advection on the decay of correlators. Specifically, the increase in wavenumber caused by advection allows density correlations to be relaxed by much smaller Brownian motions of the colloids than without advection. In this second sense, ``wavevector advection'' represents an actual physical mechanism at work in the shear-induced destruction of the colloidal glass.}

\section{Integration through transients approach}
\label{ITT}

The expressions collected in Section~\ref{model}  for the steady-state properties of dispersions under shear flow require the
stationary distribution function $\Psi_s$. This satisfies \gl{b2},
$\smop \Psi_s=0$; in this equation, time does not enter, and the shear rate $\gd$
enters only linearly, via the definition, \gl{b1}, of the SO. We anticipate however that
$\Psi_s$ does not in general depend smoothly on control parameters, such as shear rate and density or temperature.  In particular the ratio $\langle
\sigma_{xy}\rangle^{(\gd)}/\gd V \equiv \eta(\gd)$, computed via $\Psi_s$, defines the zero-shear viscosity $\eta_0$ through the limit $\eta_0\equiv\lim_{\gd\to0}\eta(\gd)$. We expect $\eta_0$ to diverge on approach to an ideal glass transition, and remain infinite
throughout the (ideal) glass phase. (If the glass transition is not ideal but
rounded by ergodicity-restoring processes, then, nevertheless, a
rapid variation of $\eta_0$ takes place.) Thus, in the ideal glass limit addressed in this paper, $\Psi_s$
is nonanalytic in $\gd$, and can only sensibly be approximated if
the glass transition mechanism is taken into account. This mechanism is kinetic, not thermodynamic, and consists
in the  arrest of the structural relaxation at high densities by `caging' and related effects. The ITT formalism for the
nonlinear rheology addresses this by calculating $\Psi_s$
via the transient dynamics. The time-dependent approach of $\Psi(t)$
to the stationary distribution at long times is found from \gl{b1},
and then approximated by considering the
slow structural rearrangements of local particle densities, along lines that follow closely those developed in the MCT of quiescent glasses.

\subsection{Generalized Green-Kubo relations}
\label{green}

As detailed in the Introduction, we approach the steady state from a shear startup protocol initialized from a Boltzmann equilibrium state at time zero. That is, the system has $\Psi(\Gamma,t=0)=\Psi_e(\Gamma)$ at
times $t\le0$; at $t=0$, a constant shear rate $\gd$ is instantaneously switched on and held constant thereafter:
\beq{c1}
\smop(\Gamma,t)
= \left\{ \begin{array}{ll}
\Omega_e(\Gamma) & t\le 0\; ,\\
\smop(\Gamma) & t> 0\; .
\end{array}\right.
\eeq
The formal solution of \gls{b1}{c1} for $t\ge0$ is: \beq{c2}
\Psi(\Gamma,t) = e^{\smop(\Gamma)\, t}\, \Psi_e(\Gamma) \; . \eeq We
assume that as $t\to\infty$, $\Psi(\Gamma,t)$ converges to a
stationary nonequilibrium distribution $\Psi_s(\Gamma)$, whose
solution from \gl{c2} can be simplified using the operator identity
\beq{} e^{\smop t}=1+\int^t_0dt' e^{\smop t'} \smop\; .\eeq
Combining this with \gl{b6} gives \rev{ for an arbitrary stationary average} \citep{Fuchs2002c,Fuchs2005a}
\rev{\beqa{c3a}\langle f \rangle^{(\gd)} &=& \int\!\!\! d\Gamma \; \Psi_s(\Gamma)\; f(\Gamma) = \int\!\!\! d\Gamma \;\Psi_e(\Gamma) \;f(\Gamma)  + \gd \;
\int_0^\infty\!\!\!\! dt \; \int\!\!\! d\Gamma\; f(\Gamma) \;e^{\smop(\Gamma)\, t }\; \Psi_e(\Gamma) \; \sigma_{xy} \nonumber\\ &=&  \langle f \rangle + \gd \;
\int_0^\infty\!\!\!\! dt \; \langle \,\sigma_{xy}\; e^{\smopb\, t}\; f\, \rangle
 \; . \eeqa}Here, the adjoint SO $\smopb$ is
introduced via partial integration. The resulting time evolution
operator now acts, to the right, on the variable $f$ whose
nonequilibrium average we want to know, rather than on the
probability density. The latter becomes time-independent and is
therefore given by the initial, Boltzmann, distribution. This formal
procedure is closely analagous to the passage from Schr\"odinger to
Heisenberg representation in quantum mechanics \citep{qmbook}. \rev{
In \gl{c3a}, the difference in the stationary and the equilibrium
average is determined from integrating up the transient correlations
between $f$, the variable of interest, and the flow induced shear
stress  fluctuation. Taking away the test function $f$ leads to the
formal expression for the stationary distribution function} \beq{c3}
\Psi_s(\Gamma) = \Psi_e(\Gamma) + \gd \; \int_0^\infty\!\!\!\! dt \;
\Psi_e(\Gamma) \; \sigma_{xy} \; e^{\smopb(\Gamma)\, t } \; . \eeq
 The
resulting \gl{c3} is central to our approach as it connects steady
state properties to time integrals formed with the shear-dependent
dynamics. Knowledge about slow relaxation processes in the system
can enter. Moreover, averages over the {\em a priori} unknown
$\Psi_s$ are converted into averages (albeit of more complicated
objects) over the Boltzmann distribution, rendering them amenable to
approximation methods developed for equilibrium dynamics.

The stationary probability distribution should, like the equilibrium one, be
normalized to unity. Therefore we require
$$0 \stackrel{!}{=} \int_0^\infty\!\!\!\! dt  \int d\Gamma\;
\Psi_e(\Gamma) \; \sigma_{xy} \; e^{\smopb(\Gamma) t }  =
\int_0^\infty\!\!\!\! dt\int d\Gamma \; e^{\smop(\Gamma) t }\;
\Psi_e(\Gamma) \; \sigma_{xy}  = \int_0^\infty\!\!\!\! dt \; \langle
\sigma_{xy} \; e^{\smopb t }\; 1\; \rangle \; .$$
Here and below, $\langle\ldots\rangle$ (without superscript) represents an {\em equilibrium} average; therefore for any constant $c$ \beq{c35} \langle
\sigma_{xy} \; e^{\smopb t }\; c\; \rangle = c\; \langle \sigma_{xy}
\rangle =0 \; ,\eeq  since the mean shear stress vanishes in
equilibrium.
The ITT expression of \gl{c3} is thus confirmed to obey the
normalization condition.

Application of \gl{c3}\rev{, or of the more explicit \gl{c3a},} is
potentially obstructed by the existence of conservation laws, which
may cause a zero eigenvalue of the (adjoint) SO, $\smopb$. The time
integration in \gls{c3}{c3a} would then not converge at long times.
This possible obstacle when performing memory function integrals,
and how to overcome it,  is familiar from equilibrium Green-Kubo
relations \citep{Forster1975}. \rev{One needs to show that the
conserved variables, which are the eigenfunctions of $\smopb$ with
zero eigenvalue, do not cause a non-decaying contribution in the
transient correlation function in \gl{c3a}. This is achieved by
considering the couplings to the densities of the conserved
quantities,  which are called `projections', as introduced by
Zwanzig and Mori and others and described by \citet{Forster1975}.}
For Brownian particles, the \remrev{density $\rho$}\rev{particle
number} is conserved. Yet, as carefully discussed by
\citet{Fuchs2005a}, density fluctuations do not couple in linear
order to the shear-induced change of the distribution function. This
follows from the vanishing of the (equilibrium) average $\langle
\sigma_{xy}\, e^{\smopb t }\, \rho_\qb \rangle =0$, and means that
no zero eigenvalue arises \add{due to \rev{a} conservation law}
\citep{Fuchs2005a}.\disc{[CONFIRM], yes}
\remrev{Moreover,}\rev{Because the projection on density
fluctuations vanishes, the orthogonal or complementary} projector
$Q$ can now be introduced \beq{c37} Q = 1 - P\; ,\quad\mbox{with }\;
P= \sum_\qb  \delta \rho_\qb \, \rangle \;\frac{1}{N S_q} \; \langle
\, \delta \rho^*_\qb\; , \eeq \add{which satisfies $Q^2=Q$, $P^2=P$
and $Q P =0$, and} in terms of which \beq{c37a} \langle \sigma_{xy}
e^{\smopb t }X \rangle = \langle \sigma_{xy} Q e^{\smopb t } Q X
\rangle = \langle \sigma_{xy} Q e^{Q\smopb Q t } Q X \rangle\; .
\eeq This holds for the observables $X = f,g$, etc. used in Section
\ref{model} to construct steady-state averages, structure functions,
and correlators. The projection step is exact, and also formally
redundant at this stage; but it will prove invaluable later on, when
approximations are performed.\rem{ Because the stress fluctuation in
\gl{c37a} lives at vanishing wavevector, a possible contribution to
the quantity $X$ arising from the non-fluctuating density needs to
be considered. Because of $\langle \sigma_{xy}\, e^{\smopb t }\,
\rho_\qb \rangle =0$, it must vanish exactly, and therefore must not
arise during approximations. To ensure this, the projection in $Q$
is chosen to contain $\rho_\qb$, including the average $\langle
\rho_\qb \rangle = \langle \rho_\qb \rangle^{(\gd)} = N
\delta_{\qb,0}$. While this subtlety in the definition of $Q$ at
$\qb=0$ does not matter when approximations are performed in
wavevector space integrations, and thus is not discussed in standard
MCT \citep{Goe:91}, it affects the results based on \gl{c37a}.}\rem{
From now on, in order to simplify the notation, density fluctuations
$\delta\rho_\qb$ will be denoted by $\rho_\qb$, dropping the
$\delta$. (Only for $\qb = 0$ is the average nonzero: $\langle
\rho_\qb \rangle = \langle \rho_\qb \rangle^{(\gd)} = N
\delta_{\qb,0}$. This quantity will be discussed separately, as
needed.)}

From the integration of the distribution function through the
transients in \gl{c3}, we gain explicit expressions for the steady-state averages in \gl{b12} (where, recall, the definition $f(\gd) \equiv \langle f_{\bf q=0} \rangle^{(\gd)} / V$ was made):
\beqa{c4} f(\gd) & = & \langle f_{\bf q=0} \rangle / V +
\frac{\gd}{V}  \; \int_0^{\infty}\!\!\!\!\! dt \; \langle
\sigma_{xy}  \, Q\; e^{Q\, \smopb \, Q \, t  }\, Q\;  \Delta f_{\bf
q=0} \, \rangle \; , \eeqa while corresponding expressions hold for
the structure functions from \gl{b13}, \beqa{c5} S_{fg; {\bf
q}}(\gd) & = &  \langle \delta f^*_{\bf q}\; \delta g_{\bf q} \,
\rangle / N + \frac{\gd}{N}  \; \int_0^{\infty} \!\!\!\!\! dt \;
\langle \sigma_{xy}\, Q  \; e^{Q\, \smopb \, Q \, t  }\, Q \;
\Delta\left(\delta f^*_{\bf q}\; \delta g_{\bf q}\right) \, \rangle
\; , \eeqa and for the fluctuation functions from \gl{b14},
\beqa{c6} C_{fg; {\bf q}}(t,\gd) &  = & \langle \delta f^*_{{\qb}}
\; e^{\smopb\, t }\, \delta g_{\qb(t)} \rangle / N  + \frac{\gd}{N}
\int_0^\infty\!\!\!\!\! dt' \; \langle  \sigma_{xy}\, Q \; e^{Q\,
\smopb\,Q\, t' }\, Q\; \Delta\left(  \delta f^*_{{\qb}}\;
 e^{\smopb\, t }\, \delta g_{\qb(t)} \right)\, \rangle \, .
\eeqa In \glto{c4}{c6}, the symbol $\Delta X$ for the fluctuation in
$X$ was introduced, \beq{c7} \Delta X_{\bf } = X_{\bf } - \langle
X_{\bf } \rangle \quad ,\qquad\mbox{thus, e.g., }\;
\Delta\left(\delta f^*_{\bf q}\; \delta g_{\bf q}\right) = \delta
f^*_{\bf q}\; \delta g_{\bf q} - N S_{fg;\qb} \; .\eeq This makes
explicit the fact that, owing to \gl{c35}, all mean values (which
are constants, for these purposes) drop out of the ITT integrals,
leaving only the fluctuating parts to contribute. Note also that all
the averages, denoted $\langle\ldots\rangle,$ throughout
\glto{c4}{c6} are evaluated within the (Boltzmann) equilibrium
distribution $\Psi_e(\Gamma)$. These manipulations may be
unfamiliar, but have antecedents in the literature; when studying
the nonlinear rheology of simple fluids far from any glass
transition,\add{ generalized Green-Kubo relations based on}
transient correlation functions\rem{ related to}\add{ like \rev{those} in}
\gl{c4} \disc{[DO YOU MEAN \gl{c5}]; clearer?} were found useful in
thermostated simulations \citep{Morriss87} and in mode coupling
approaches \citep{Kawasaki1973b}.

\subsection{Transient density correlator}
\label{trade}

 The problem of calculating steady state averages shall next
be converted into one of first finding the transient response after
startup of steady shear, and then integrating this response in order
to use \glto{c4}{c6}. To keep track of shear-induced structural
rearrangements we define the following transient density correlator
at wavevector $\qb$: \beq{d1} \Phi_\qb(t) = \frac{1}{N S_q}\;
\langle \delta \rho_{{\qb}}^* \;
 e^{\smopb t}\;  \delta \rho_{\qb(t)} \rangle
\eeq which differs \remrev{somewhat} from the choice made by
\citet{Fuchs2002c} as detailed in Appendix B. Note that
$\Phi_\qb(t)$ depends on the shear rate $\dot\gamma$; for notational
brevity, this argument is suppressed. Its interpretation follows
from the joint probability $\bar W_2(\Gamma t,\Gamma' t')$ that the
system \rev{{\em in equilibrium} was} at state point $\Gamma'$ at time
$t'$, and that shear and internal dynamics then take the system to
state point $\Gamma$ at time $t>t'$. Denoting as $\bar P(\Gamma
t|\Gamma' t')$ the conditional probability of evolving via the SO,
$\smop$, in \gl{b1} from state $\Gamma'$ to state $\Gamma$, the
desired joint probability is given by
$$\bar W_2(\Gamma t,\Gamma' t')= \bar P(\Gamma t|\Gamma' t')\, \Psi_e(\Gamma') = e^{\smop(\Gamma)\, (t-t')} \; \delta(\Gamma-\Gamma') \; \Psi_e(\Gamma')\; .$$
To derive this, we used the fact that the transition probability
$\bar P$ also obeys the Smoluchowski equation (\ref{b1}), and that
its formal solution can be given using the initial condition that
$\Gamma$ and $\Gamma'$ coincide: $\bar P(\Gamma t|\Gamma'
t)=\delta(\Gamma-\Gamma')$. The transient correlator is now defined
as the stochastic overlap between an equilibrium density fluctuation
$\delta \rho^*_{\qb}$ with wavevector $\qb$ at time $t'=0$ and an
appropriately shear-advected density fluctuation $\delta
\rho_{\qb(t)}$ at later time $t$, where the time evolution between
these times is given by the full SO (and so allows for the presence
of shear, interactions, and diffusion): \beqa{d2}
\Phi_\qb(t) &=& \frac{1}{N S_q}\; \int\!\!\!d\Gamma\!\int\!\!\!d\Gamma'\; \bar W_2(\Gamma t,\Gamma' 0) \;
\delta\rho^*_\qb(\Gamma') \;  \delta\rho_{\qb(t)}(\Gamma)\nonumber\\
&=& \frac{1}{N S_q}\; \int\!\!\!d\Gamma\!\int\!\!\!d\Gamma'\;
\delta\rho_{\qb(t)}(\Gamma)\; e^{\smop(\Gamma)\, (t-t')} \;
\delta(\Gamma-\Gamma') \; \Psi_e(\Gamma')\;
\delta\rho^*_\qb(\Gamma')\; .\nonumber \eeqa Partial integrations,
which bring in the adjoint SO $\smopb$, followed by integration over
$\Gamma'$, then lead directly to \gl{d1}.

From its definition and inversion symmetry, it follows that our
correlator is real and symmetric in $\qb$:
$\Phi^*_\qb(t)=\Phi_\qb(t)=\Phi_{-\qb}(t)$. In the absence of both
Brownian motion \add{($D_0=0$)} and particle interactions, this
correlator does not decay at all ($\Phi_\qb(t)=1$); a density
fluctuation which is solely advected is tracked perfectly by the
wavevector advection that was built into the correlator definition,
\gl{d1}.\rem{ Moreover, in the absence of Brownian motion ($D_0=0$)
the mobility $D_0k_BT$ also vanishes so that for any finite
interaction potential, the correlator again does not decay. However,
when hard-core (excluded volume) interactions are present, the limit
of infinite bare Peclet number, Pe$_0$) should be singular because
of hard-core particle collisions \citep{Brady1997}. For finite
$D_0$, which is the only case considered here, our correlator should
always decay initially at short times and in any ergodic state will
eventually decay to zero. [I HAVE NO IDEA WHAT THIS PARAGRAPH IS
DOING HERE!]}

We note in passing that the transition probability $\bar P(\Gamma
t|\Gamma' t')$ (which was used not only to obtain \gl{d1} but also
in the corresponding definition of the time-dependent correlators in
\gl{b14}) is connected to the time-dependent distribution function
$\Psi(\Gamma,t)$ via the relation (for $t>t'>0$):
$$ \Psi(\Gamma,t) = \int d\Gamma'\; \bar P(\Gamma t|\Gamma' t') \;  \Psi(\Gamma',t') =
\int d\Gamma'\; e^{\smop(\Gamma)\, (t-t')} \; \delta(\Gamma-\Gamma') e^{\smop(\Gamma')\, t'} \; \Psi_e(\Gamma')$$
$$ =
 e^{\smop(\Gamma)\, (t-t')} \;  e^{\smop(\Gamma)\, t'} \; \Psi_e(\Gamma)
 =
 e^{\smop(\Gamma)\, t} \; \Psi_e(\Gamma)\; .$$
The probability of occupying state point $\Gamma$ at time $t$ is
thus given by the product of the probability of being at $\Gamma'$
at earlier time $t'$, and the transition probability during the time
interval $t-t'$, integrated over $\Gamma'$. Reassuringly, the formal
manipulations used within ITT conserve the consistency between one-
and two-time probabilities required by the Chapman-Kolmogorov
relations for Markovian processes \citep{vankampen}.

\subsection{Coupling to structural relaxation}
\label{coupling}

The ITT expressions, \glto{c4}{c6}, leave us with the problem of how
to approximate time-dependent correlation functions of the form
$\langle \sigma_{xy}\;Q\, e^{Q\smopb Q\, t}\, Q\; \Delta X \rangle$.
Here, as before, $\Delta X$ denotes a general fluctuation. (In the
case of the temporal correlators $C_{fg; {\bf q}}(t',\gd)$ in
\gl{c6}, $\Delta X$ depends on another time, $t'$.) The
interpretation of these  functions can be learned from their
definition \beq{d3} \langle \sigma_{xy}\; Qe^{Q\smopb Q\, t}\, Q
\Delta X \rangle = \int\!\!\!d\Gamma\!\int\!\!\!d\Gamma'\; \bar
W^Q_2(\Gamma t,\Gamma' 0) \; \sigma_{xy}(\Gamma') \;  \Delta
X(\Gamma)\; . \eeq \disc{[MUST ADD EQUATION HERE EXPLICITLY DEFINING
$\bar W^Q_2$.]}\add{ Here, the joint probability $\bar W^Q_2(\Gamma
t,\Gamma' 0)$ is formed identically to $\bar W_2(\Gamma t,\Gamma'
0)$ in the previous Sect. \ref{trade}, \mike{except} that
the time evolution is given by the reduced SO $Q\smop Q$, and that
the overlap of fluctuations with density at \mike{both} initial and \mike{final}
times is eliminated \rev{(again using the projector $Q$)}.} At time zero, an equilibrium stress
fluctuation arises; the system then evolves under internal and
shear-driven motion until time $t$, when its correlation with a
fluctuation $\Delta X$ is determined. Integrating up these
contributions for all times since the start of shearing gives the
difference \rev{between} the shear-dependent quantities \rev{and} the equilibrium
ones. During the considered time evolution, the projector $Q$
prevents linear couplings to the conserved particle density from
arising. As stated previously, this projection is optional within
the current (exact) manipulations since \add{no such} couplings
arise within the exact dynamics. But in any approximate dynamics
they might arise (in which case their removal by projection is
required, to avoid artefacts).

The time dependence and magnitudes of the correlations in \gl{d3}
shall now be approximated by using the overlaps of both the stress
and $\Delta X$ fluctuations with appropriately chosen {\em `relevant
slow fluctuations}'. For the dense colloidal dispersions of
interest, the relevant structural rearrangements will be described
as usual in terms of {\em density fluctuations}. Because of the
projector $Q$ in \glto{c4}{c6}, the lowest nonzero order in
fluctuation amplitudes, which we presume dominant, must then involve
pair-products of density fluctuations, $\rho_\kb\,\rho_\pb$. (These
are familiar \mike{elements} in the MCT for the quiescent glass transition.)
In accord with the interpretation of \gl{d3}, we choose to take a
static (equal-time) overlap between the fluctuation $\Delta X$ and
such density pairs, whose time evolution in relation to the earlier
stress fluctuation is then approximated using the transient density
correlator defined in \gl{d1}.

\remrev{More generally,}\rev{To present the general case,} we give here the mode coupling approximation for the
time-dependent transient correlator of any two variables, $f$ and
$g$ (which may themselves be composite quantities) that do not
couple linearly to density, and that evolve in time according to the
projected (or `reduced') SO, $Q\smopb Q$. First, the example
$\langle \delta f^*_{\qb(-t)}\; Q \; e^{Q\, \smopb\, Q\, t}\; Q \;
\delta g_{\qb} \rangle$ is considered. A projector onto density
pairs is introduced as \beq{d35} P_2 = \sum_{\kb > \pb} \delta
\rho_\kb\,\delta\rho_\pb \rangle \; \frac{1}{N^2 S_k S_p} \; \langle
\delta\rho^*_\kb \,\delta\rho^*_\pb \; , \eeq where the Gaussian
approximation, $ \langle \delta\rho^*_\kb \,\delta\rho^*_\pb\,
\delta\rho_{\kb'}\,\delta\rho_{\pb'} \rangle \approx N^2\, S_k \,S_p
\; \delta_{\kb,\kb'}\, \delta_{\pb,\pb'}$, was used to simplify the
denominator. (This is again standard \citep{Goe:91}.) The ordering
$\kb > \pb$ and $\kb' > \pb'$ should be kept in mind. Fluctuations
are approximated by their overlap with density pairs, as follows:
$$\langle \delta f^*_{\qb(-t)}\; Q \; e^{Q\, \smopb\, Q\, t}\; Q \; \delta g_{\qb} \rangle \approx
\sum_{\kb > \pb}\sum_{\kb' > \pb'} \frac{ \langle \delta
f^*_{\qb(-t)}\, Q \;\delta\rho_{\kb'}\,\delta\rho_{\pb'} \rangle
}{N^2 S_{k'} S_{p'}} \; \langle \delta\rho^*_{\kb'}
\,\delta\rho^*_{\pb'}\; e^{Q\, \smopb\, Q\, t}\;
\delta\rho_{\kb}\,\delta\rho_{\pb} \rangle \; \frac{ \langle
\delta\rho^*_{\kb}\,\delta\rho^*_{\pb}\; Q\, \delta g_{\qb} \rangle
}{N^2 S_{k} S_{p}}\;
$$
where the last term vanishes unless $\qb=\kb+\pb$. A crucial step in
the mode coupling theory is now to (a) break the four-density
average into a product of pair averages; and (b) replace the reduced
dynamics with the full one \citep{Kawasaki,Goe:92}: \beq{d36}
\langle \delta\rho^*_{\kb'} \,\delta\rho^*_{\pb'}\; e^{Q\, \smopb\,
Q\, t}\; \delta\rho_{\kb}\,\delta\rho_{\pb} \rangle
 \approx
N^2\, S_{k(-t)}\, S_{p(-t)}\; \Phi_{\kb(-t)}(t)\;
\Phi_{\pb(-t)}(t)\; \delta_{\kb(-t),\kb'}\,\delta_{\pb(-t),\pb'}\; .
\eeq \rev{The $S_{k(-t)}$ are the equilibrium structure factors
evaluated with the (magnitude of the) time-dependent wavevector $\kb(-t)=\kb ( 1 + \kap\, t)$, and capture the affine stretching of equilibrium density
fluctuations; see Appendix B for their different handling here and
by \citet{Fuchs2005a}.} \disc{[EXPLAIN WHY IS (b) NECESSARY --
COULDN'T WE EQUALLY WELL START FROM EXPRESSIONS WITHOUT Q's INSIDE
THE EXPONENTIAL?] As $Q$ was not required in the exact expression, I
would claim that it doesn't make a difference here. But, the
'philosophy' of the factorisation in \gl{d36} is that only the
dynamics in the part of Hilbert-space not containing linear density
fluctuations is approximated in this fashion. The philosophy would
suggest, to first use projection-operator steps so that couplings to
linear densities are explicitly given (thereby introducing the $Q$'s
into the SO), and that the factorisation is used only for the rest.
This prescription actually goes back to Kawasaki who considered the
hydrodynamic and order parameter fields. Goetze adopted  this
prescription, also.} Collecting all terms, and enforcing the
wavevector restrictions following from translational invariance, we
obtain
$$\langle \delta f^*_{\qb(-t)}\; Q \; e^{Q\, \smopb\, Q\, t}\; Q \; \delta g_{\qb} \rangle \approx
\sum_{\stackrel{\kb > \pb}{\kb+\pb=\qb}} \frac{ \langle \delta
f^*_{\qb(-t)}\, Q \;\delta\rho_{\kb(-t)}\,\delta\rho_{\pb(-t)}
\rangle\; \langle \delta\rho^*_{\kb}\,\delta\rho^*_{\pb}\; Q\,
\delta g_{\qb} \rangle }{N^2 S_{k} S_{p}} \; \Phi_{\kb(-t)}(t)\;
\Phi_{\pb(-t)}(t)\;
$$
$$\phantom{\langle \delta f^*_{\qb(-t)}\; Q \; e^{Q\, \smopb\, Q\, t}\; Q \; \delta g_{\qb} \rangle} =
\sum_{\stackrel{\kb' > \pb'}{\kb'+\pb'=\qb(-t)}} \frac{ \langle
\delta f^*_{\qb(-t)}\, Q \;\delta\rho_{\kb'}\,\delta\rho_{\pb'}
\rangle\; \langle \delta\rho^*_{\kb'(t)}\,\delta\rho^*_{\pb'(t)}\;
Q\, \delta g_{\qb} \rangle }{N^2 S_{k'(t)} S_{p'(t)}} \;
\Phi_{\kb'}(t)\; \Phi_{\pb'}(t)\; .
$$
The last equality follows from a change of dummy-summation indices
from $\kb$ to $\kb'=\kb(-t)$ and  from $\pb$ to $\pb'=\pb(-t)$; the
other wavevectors  are shifted from $\kb$ to $\kb'(t)$, etc.. A
similar shift can be performed in a number of analagous summations
discussed below, but for brevity this will not be notated explicitly
in each case.

The above mode coupling procedure can be summarized as a rule that applies to all fluctuation products that exhibit slow structural relaxations but whose variables cannot couple linearly to the density:
\beq{d5} Q\; e^{Q \smopb Q t}\; Q \approx \sum_{
\kb > \pb}\;  Q\, \delta\varrho_{{\kb(-t)}}\,
\delta\varrho_{{\pb(-t)}}\; \rangle \; \frac{\Phi_{\kb(-t)}(t) \;
\Phi_{\pb(-t)}(t)}{N^2 S_{{ k}}\,  S_{{ p}}}\; \langle
\;\delta\varrho^*_{\kb} \, \delta\varrho^*_{\pb}\, Q \eeq The
fluctuating variables are thereby projected onto pair-density
fluctuations, and the time-dependence follows from that of the
transient density correlators $\Phi_{\qb(-t)}(t)$. These describe
the relaxation (caused by shear, interactions and Brownian motion)
of density fluctuations with equilibrium amplitudes.  Higher order
density averages are factorized into products of these correlators,
and the reduced dynamics containing the projector $Q$ is replaced by
the full dynamics. The entire procedure is written in terms of {\em
equilibrium} averages, which can then be used to compute
nonequilibrium steady states via the ITT procedure.

A second rule is needed for fluctuations that can couple linearly to
densities. This is derived in complete analogy to the case of
pair-densities just discussed: \beq{d4} e^{\smopb t} \approx
\sum_{\bf q}\;   \delta\varrho_{{\qb(-t)}} \; \rangle \;
\frac{\Phi_{\qb(-t)}(t)}{N S_{{ q}}} \; \langle \;
\delta\varrho^*_{{\bf q}} \eeq As a check of Eq. (\ref{d4}), we note
that it is consistent with the definition of the transient
correlator as can be seen from
$$ N S_q\; \Phi_\qb(t) = \langle \delta\rho^*_\qb \; e^{\smopb t}\; \delta\rho_{\qb(t)} \rangle \approx
\sum_{\qb'}\;  \langle \delta\rho^*_\qb \;
\delta\varrho_{{\qb'(-t)}} \; \rangle \; \frac{\Phi_{\qb'(-t)}(t)}{N
S_{{\qb'}}} \; \langle \; \delta\varrho^*_{{\qb'}}\;
\delta\rho_{\qb(t)} \rangle
$$ $$ = \sum_{\qb'}\; N S_{q}\; \delta_{\qb,\qb'(-t)}\;
\frac{\Phi_{\qb}(t)}{N S_{\qb(t)}}\; N S_{\qb(t)}\;
\delta_{\qb',\qb(t)} = N S_q\; \Phi_\qb(t)
$$
where the Kronecker-$\delta$'s lead to the expected result because
$\qb'(-t)=\qb' \cdot (1+ \kap t)=\qb(t) \cdot (1+ \kap t) = \qb$. A similar consistency
check can be applied to \gl{d5} for the pair-product density
fluctuations calculated in detail above.

\subsubsection{Steady state averages}

The mode coupling approximations introduced above can now be applied
to the exact generalized Green-Kubo relations from
Sect.~\ref{green}. Steady state expectation values from \gl{c4}, for
variables that do not couple linearly to density fluctuations, are
approximated by projection onto pair density modes, giving by the
first rule discussed above \beq{d6} f(\gd) \approx \langle f_{\bf 0}
\rangle / V + \frac{\gd}{2V} \int_0^\infty\!\!\!dt\; \sum_{\kb}
\frac{k_xk_y(-t)S'_{k(-t)}}{k(-t)\; S^2_{k}}\; V^{f}_{\kb} \;
 \Phi^2_{\kb(-t)}(t)\, \; ,
\eeq with $t$ the time since switch-on of shear. To derive this, the
property  $\Phi^*_\kb=\Phi_{-\kb}=\Phi_{\kb}$ was used; also the
restriction $\kb>\pb$ when summing over wavevectors was dropped, and
a factor $\frac 12$ introduced, in order to have unrestricted sums
over $\kb$. Within \gl{d6} we have already substituted the following
explicit result for the equilibrium correlation of the shear stress
with density products: \beq{d7} \langle \sigma_{xy} \; Q \;
\delta\rho_{\kb(-t)} \; \delta\rho_{\pb(-t)} \rangle = N \frac{k_x
\, k_y(-t)}{k(-t)}\; S'_{k(-t)} \; \delta_{\kb(-t),-\pb(-t)} =
\frac{N}{\gd} \;
\partial_t S_{k(-t)} \; \delta_{\kb,-\pb}\; , \eeq \rem{It is an
exact equality using the equilibrium distribution function and
\gl{b6}
$$\langle \sigma_{xy} \; Q \;
\delta\rho_{\kb} \; \delta\rho^*_{\kb} \rangle = \langle \sigma_{xy}
\; \delta\rho_{\kb} \; \delta\rho^*_{\kb} \rangle = \int d\Gamma\;
\Psi_e (-\sum_i F^x_i y_i) \; \delta\rho_{\kb} \; \delta\rho^*_{\kb}
= \int d\Gamma \; \Psi_e (\sum_i \partial^x_i y_i)\;
\delta\rho_{\kb} \; \delta \rho^*_{\kb}
$$
$$\phantom{\langle \sigma_{xy} \; Q \; \rho_{\kb} \; \rho^*_{\kb}
\rangle} = i k_x \sum_{ij} \langle y_i(
e^{i\kb(\rb_i-\rb_j)}-e^{-i\kb(\rb_i-\rb_j)}) \rangle  = i k_x
\frac{\partial}{\partial k_y} \sum_{ij} \langle
e^{i\kb(\rb_i-\rb_j)} \rangle \;. $$}\disc{Mike, should we show the
explicit calculation? here or in the appendix?}\add{which can be
calculated as in equilibrium \citep{Goe:91} \mike{using $\Psi_e$ and
\gl{b6},} considering time $t$ in the advected wavevectors as fixed
parameter.} The wavevector derivative appears \mike{as}
$S'_k\equiv\partial S_k/\partial k$; the second equality, involving
the time derivative, is useful and will be discussed further in
Section \ref{disc}. Equation (\ref{d6}), as derived via the
mode-coupling rule detailed above, contains a `vertex function'
$V^f_{\kb}$, describing the coupling of the desired variable $f$ to
density pairs. This denotes the following quantity \beq{d8}
V^{f}_{\kb}\equiv \langle \delta\rho_{\kb}^*\delta\rho_{\kb} \; Q \;
\Delta f_{\bf 0} \rangle / N = \langle
\delta\rho_{\kb}^*\delta\rho_{\kb} \;\Delta f_{\bf 0} \rangle / N -
S_0 \left( S_k + n \, \frac{\partial S_k}{\partial n} \right)
\left.\frac{\partial f^\text{eq}}{\partial n} \right)_T\; , \eeq To
obtain the second result, two thermodynamic equalities were used:
the first is \citet{baxter}'s relation $\langle\delta\rho_{\kb}^*\,
\delta\rho_{\kb}\, \delta\rho_{\bf 0} \rangle = N S_0 \left( S_k + n
\, \frac{\partial S_k}{\partial n} \right)$, the second is a result
for the thermodynamic derivative, $\langle \delta \rho_{\bf q}^*\,
\Delta f_{\bf q} \rangle/ \langle |\delta \rho_\qb|^2\rangle \to
\frac{\partial f^\text{eq}}{\partial n} )_T$ for $q\to 0$ , in which
$f^\text{eq}=\langle f_{\bf 0} \rangle / V$ is viewed as an
(intensive) thermodynamic density \citep{Forster1975,GoetzeLatz}.
\mike{The calculation of the term in $S_0$ is technically somewhat
delicate; while this term has no effect on the calculation of
rheological properties such as stress, it does influence the
distorted structure factor addressed in the next Section.}

\disc{Mike, I am now convinced that the use of Baxter's relation
does not require caution about the differences between $\qb=0$ and
the limit $q\to0$, because: \\ Baxter considers a canonical system
at fixed $N$. He derives (I did not check his derivation) the
relation between the total (indirect) correlation (cluster)
functions $h_{12}(\rb,\rb')$ and $h_{123}(\rb,\rb',\rb'')$ (with
$S_0$ connected to the compressibility)
$$(n S_0 \partial_n - 2 )\;
h_{12}(\rb,\rb') = \int d\rb''\, h_{123}(\rb,\rb',\rb'')$$ These
functions fall off at large distances ${\bf s}=\rb-\rb'$ and ${\bf
s}'=\rb''-\rb'$ rapidly in homogeneous states, so that they can be
Fourier-transformed
$$
\hat h_{12}(\kb) = \int d{\bf s}\; e^{i \kb \cdot {\bf s}}
\;h_{12}({\bf s})= \frac 1V \int d\rb d\rb'\; e^{i \kb \cdot
(\rb-\rb')} \;h_{12}(\rb,\rb')
$$
$$
\hat h_{123}(\kb,\kb') = \int d{\bf s}d{\bf s}'\; e^{i \kb \cdot
{\bf s}}\; e^{i \kb' \cdot {\bf s}'} \;h_{123}({\bf s},{\bf s}')=
\frac 1V \int d\rb d\rb' d\rb''\; e^{i \kb \cdot (\rb-\rb')}\; e^{i
\kb' \cdot (\rb''-\rb')} \;h_{123}(\rb,\rb',\rb'')
$$
so that  the limit $k\to0$ is analytic in them. Baxter's relation
becomes:
$$(n S_0 \partial_n - 2 )\;
\hat h_{12}(\kb) = \hat h_{123}(\kb,0)$$ if one interprets the
derivative $\partial_n$ to be $\partial_n=V \partial_N$ at fixed
volume, to commute differentiation and (Fourier-) integration.
 Converting
them into (our) structure factors requires to consider the
self-correlations, which are not contained in them (!). Defining the
structure factors using the density fluctuations:
$$S(k)= \frac 1N \langle |\delta \rho_\kb |^2 \rangle \qquad
\text{and }\quad S(\kb,\kb')= \frac 1N \langle \delta \rho_\kb\;
\delta \rho_{\kb'} \; \delta \rho^*_{\kb+\kb'} \rangle$$ so that no
(singular) contributions at $\kb=0$ appear in the structure factors,
the connection follows when taking the self-terms into account:
$$\hat h_{12}(\kb) = n ( S(\kb) - 1) $$ and
$$\hat h_{123}(\kb,\kb') = n \left(  S(\kb,\kb') - S(k) - S(k')
- S(|\kb+\kb'|) + 2 \right)$$ Putting this together gives the
relation we use:
$$\frac 1N\; \langle \delta \rho_\kb\;
\delta \rho_{0} \; \delta \rho^*_{\kb} \rangle = S(k,0) = S_0
\partial_n\; \frac 1V\; \langle |\delta \rho_\kb |^2 \rangle = S_0
\left( S(k) + n
\partial_n S(k)\right)$$
This also suggests, where the previous inconsistency between
Baxter's relation and $\langle \delta \rho_{\bf 0}^*\, \Delta f_{\bf
0} \rangle = n S_0 \, \frac{\partial \langle f_{\bf 0}
\rangle}{\partial n} )_T$  comes from. The latter is wrong, and the
corrected version needs to be as given now in the text. It is now
consistent with Baxter if the quantity is $f_{\bf 0}=|\delta\rho_\qb|^2$.\\
This leaves the problem, however, that this is not what we want for
calculating the vertex in \gl{d12}, because it leads to problems at
large $q$. They arise from the self-correlations of the individual
particles in $S_q$ at $q\to\infty$, which, I believe, we should
not approximate by coupling to density pairs. \\
btw, there follows a surprising prediction from Baxter's relation:
$$S(k\to\infty , 0) = S_0$$ Have you heard of this?
 }\mikecom{This all seems much more consistent now. It is slightly
annoying to conclude that MCT fails at high $q$ but I see no good
reason to expect it to work there, i.e. the fact that it is OK for
the quiescent case may be mainly good luck. I've not heard of the
result you ask about.}

The general result, \gl{d6}, can now be applied to compute any
stationary expectation value, including for example the shear
stress, $\langle\sigma_{xy}\rangle^{(\gd)} /V$. From \glto{d6}{d8}
one finds the vertex $V^{\sigma_{xy}}_{\kb} = (k_xk_y/k)  S'_k$,
which gives the explicit ITT approximation for  the stationary shear
stress of a homogeneously sheared dispersion \citep{Fuchs2002c}
\beq{d9} \sigma(\gd) \equiv \langle \sigma_{xy}/V \rangle^{(\gd)}
\approx \frac{k_BT \,\gd}{2}  \int_0^\infty\!\!\!\!dt\,
\int\!\!\frac{d^3k}{(2\pi)^3}\; \frac{k_x^2k_yk_y(-t)}{k\, k(-t)}\;
\frac{S'_kS'_{k(-t)}}{S^2_{k}}\; \Phi^2_{\kb(-t)}(t)\; . \eeq

\subsubsection{Structure functions}
\label{avstruc}

 The structure functions from \gl{c5} can be
approximated along identical lines \mike{to} the stationary averages
and become \beq{d10} S_{fg;\qb}(\gd) \approx \langle \delta f^*_\qb
\delta g_\qb  \rangle / N + \frac{\gd}{2N}
\int_0^\infty\!\!\!\!\!\!dt\, \sum_{\kb}
\frac{k_xk_y(-t)S'_{k(-t)}}{k(-t) S^2_{k}}\; V^{fg}_{\kb} \;
\Phi^2_{\kb(-t)}(t)\; , \eeq with the general vertex built
\mike{just} as in \gl{d8}: \beq{d11} V_{\bf
k}^{fg}=\langle\delta\rho_{\bf k}^{*}\delta\rho_{\bf k}Q\Delta
\left( \delta f^{*}_\qb\delta g_\qb \right) \rangle / N \; .\eeq
Applying this general result to the important case of the distorted
structure factor under shear requires\rem{\disc{UNFORTUNATELY WRONG:
I MISSED A TERM some preparative considerations. It would be
un-physical to approximate the term in $S_\qb(\gd)$ arising from the
single particle (self-) correlations by projection onto density
pairs. Thus the quantity to be approximated should be the
intermolecular density fluctuations $\sum_{i\ne j}
\exp{i\qb\cdot(\rb_i-\rb_j)}=|\delta\rho_\qb|^2-\rho_{\bf 0}$} at
$\qb\ne0$. The vertex can then be} \rev{use of} the vertex built with density
pairs, which can be evaluated as in \gl{d8}: \beq{d12} V_{\bf
k}^{\rho^*_{\bf q} \rho_{\bf q}}=\langle\delta\rho_{\bf
k}^{*}\delta\rho_{\bf k}\; Q\; \Delta\left( \delta\rho_{\bf
q}^{*}\delta\rho_{\bf q} \right) \rangle / N =2 N S_q^2\;
\delta_{\bf q k}-  S_0\; \left(S_k + n \frac{\partial S_k}{\partial
n}\right) \frac{\partial}{\partial n}\left(n S_q \right)\; . \eeq
The first term results from a Gaussian decoupling approximation of
four density fluctuations into products of pairs; its factor 2
follows from Wick's theorem\add{ (including symmetry)}, and cancels
the factor $\frac 12$ introduced in \gl{d6} in order to have
unrestricted sums.\add{ The second term again follows from Baxter's
relation, or alternatively the thermodynamic derivative as given
\mike{after} \gl{d8}.} This vertex leads to the ITT approximation:
\beqa{d13}
S_{\bf q}(\dot{\gamma})&=& S_q +\dot{\gamma}\,  \left\{\int_0^\infty dt\; \frac{q_x q_y(-t)}{q(-t)}S'_{q(-t)}\; \Phi^2_{\bf q(-t)}(t) \right\}\\
&-& \frac{\dot{\gamma}\, S_0}{2n}\; \left( S_q + n \frac{\partial
S_q}{\partial n} \right)\; \left\{\int_0^\infty dt
\int\frac{d^3k}{(2\pi)^3} \frac{k_x k_y(-t)}{k(-t)S^2_{k}}
S'_{k(-t)}\left( S_k + n \frac{\partial S_k}{\partial n}\right)
\Phi^2_{\bf k(-t)}(t)\right\}\nonumber, \eeqa where the second term
on the right is anisotropic, but the third is isotropic.\add{ This
result satisfies $S_{\qb}(\gd,n\to0)\to 1 + {\cal O}(n)$ but not
$S_{\qb\to\infty}(\gd)\to 1$ , as required when only the
self-correlations survive in both limits; in \gl{d13}, the isotropic
contribution does not vanish for $q\to\infty$. Apparently, the mode
coupling approximation breaks down at large wavevectors where it
\mike{cannot} properly resolve the very local correlations, and
misses \mike{the fact} that all intermolecular contributions to
$S_\qb(\gd)$ should vanish there. A previous expression for the
isotropic term published by \citet{Henrich2007} \mike{inadvertently
assumed that this limiting behaviour would be correct; for now, the
form above replaces that result as the formal prediction from
ITT/MCT. As mentioned previously, however, this term stems from the
somewhat delicate $S_0$ piece \rev{in} \gl{d8} and it is possible that an
improved treament will later be found that can restore full
consistency to this aspect of the theory. Because of the uncertainty
over the $q$ range for which this error becomes important, we do not
discuss further the isotropic term in the following. However, in
numerical solutions, \citet{Hen:07} found it to be subdominant to
the first and second term on the right hand side of \gl{d13} for all
wavevectors below the second peak in the equilibrium $S_q$.}}
\disc{Mike, this is the present (sad) status of it.}

\subsubsection{Stationary temporal correlators}

Temporal correlators $C_{fg;\qb}(t,\gd)$ from \gl{c6}, describing
the stationary, time-dependent fluctuations in the sheared system,
are next approximated. We refrain from giving the general unwieldy
expressions, but note that the derivation follows the same method as
detailed above, except that a linear coupling of $e^{\smopb t}$ to
density is possible in $C_{fg;\qb}(t,\gd)$. Whenever this coupling
does not vanish for specific reasons, the general approximation for
the $C_{fg;\qb}(t,\gd)$ follows from: $(i)$ using \gl{d4} on the
$t$-dependence in \gl{c6}, and $(ii)$ using \gl{d5} on the
$t'$-dependence there. (The former is the time delay in the
correlator, the latter is the dummy variable in the ITT integral.)
As a concrete example, we state here the resulting ITT approximation
for the stationary density correlator under shear: \beqa{d17}
C_\qb(t,\gd) &\approx& S_q\; \Phi_\qb(t) + \frac{\gd}{N}\;
\sum_{\qb'} \int_0^\infty dt'\; \langle \sigma_{xy}\, Q\; e^{Q
\smopb Q t'}\; Q \delta \rho^*_\qb\; \delta\rho_{\qb'(-t)} \rangle
\frac{\Phi_{\qb'(-t)}(t)}{NS_{q'}} \; \langle \delta\rho^*_{\qb'}\, \delta\rho_{\qb(t)} \rangle\nonumber\\
&=&  \left[ S_q + \frac{\gd}{N}  \; \int_0^{\infty} dt' \; \langle
\sigma_{xy}\, Q  \; e^{Q\, \smopb \, Q \, t'  }\, Q \;
\Delta\left(\delta\rho^*_{\qb}\; \delta\rho_{\qb} \right) \,
\rangle\right] \Phi_\qb(t) = S_\qb(\gd) \; \Phi_\qb(t)\; . \eeqa
It should be emphasized at this point that the stationary correlator
$C_{\qb}(t,\dot\gamma)$ is conceptually a quite different object
from the transient one, $\Phi_{\qb}(t)$ (whose dependence on
$\dot\gamma$ is, we recall, notationally suppressed). Specifically,
the first describes correlations between two nonequilibrium states,
separated by $t$ in time, with both states arising long after
startup and in the stationary regime. The second describes
correlations between an initial equilibrium state at time zero and a
transient nonequilibrium one at finite time $t$ after startup.
Nonetheless, within our approximation scheme, the stationary density
correlators differ from the transient ones only by a static
renormalization of the amplitude. Moreover, this renormalization
coincides with the distortion of the structure factor under shear,
which is of course the initial value $C_\qb(t=0,\gd)$. Therefore, if
both  are normalized to unity at time zero, {\em the transient and
stationary density correlators precisely coincide within our chosen
MCT approximation}. This is arguably a drawback, since discernible
differences between transient and stationary correlators at
intermediate times have been observed in both confocal microscopy
and computer simulations by \citet{Zau:08}. Very recently, an
extended approximation scheme has been suggested by \citet{Kru:08},
which gives a more faithful account of these differences. However
that extension is not immediately generalizable to other quantities,
so we do not pursue it here.

\section{Transient density fluctuations}
\label{tradens}

The transient density correlators $\Phi_\qb(t)$ are defined by \gl{d1} above, viz.,
\beq{} \Phi_\qb(t) = \frac{1}{N S_q}\; \langle
\rho_{{\qb}}^* \;
 e^{\smopb t} \rho_{\qb(t)} \rangle\;, \nonumber
\eeq \add{where, from now on, in order to simplify the notation,
density fluctuations $\delta\rho_\qb$ will be denoted by $\rho_\qb$,
dropping the $\delta$, \mike{which is anyway redundant for $\qb\ne {\bf 0}$.}}  The time
evolution operator causes decay by a combination of shear and
Brownian effects, while the wavevector advection in $\rho_{\qb(t)}$
accounts for affine particle motion. Thus the $\Phi_\qb(t)$ are
matrix-elements of the time-evolution operator $e^{\smopb t}$
sandwiched between $\rho_\qb$ on the left and $\rho_{\qb(t)}$ on the
right side, with averaging performed over the equilibrium
(Boltzmann) distribution $\Psi_e\sim e^{-U}$. The usefulness of the
ITT approach rests on the availability of good approximations to
these $\Phi_\qb(t)$, whose exact equations of motion shall be
determined next. Projection operator manipulations, following
standard procedures for systems close to equilibrium, are thereafter
employed to set up equations of motion suitable for closure by \rev{a subsequent} mode-coupling approximation. The aim is to create an
approximate description in which the slow structural dynamics, which
is certainly present in quiescent situations, can compete with
shear-driven relaxation caused by wavevector advection.

\subsection{Equations of motion}

The time-dependent advection of the wavevector for
a density fluctuation can be written using the operator $\dod$ from \gl{b4}:
\beq{e1}
\rho_{\qb(t)}\rangle = e^{-\dod t}\; \rho_\qb\rangle ,
\eeq
as can be seen easily from
$$e^{-\dod t}\; \rho_\qb\rangle = \sum_i e^{-\dod t}\; e^{i \qb\cdot\rb_i} =
\sum_i e^{- \gd t\, y_i \, \partial/\partial x_i} \;  e^{i
\qb\cdot\rb_i} = \sum_i e^{- \gd t\, y_i q_x} \;  e^{i
\qb\cdot\rb_i}\; .$$
(The `ket' symbol, $\rangle$, emphasises that, since the time evolution operator acts on everything to its right, further terms will arise if any other variables lie to the right of $\rho_\qb$.)
Thus, the operator $\dod$ is identified as
causing affine advection in density fluctuations. It possesses this
interpretation also for higher powers of density fluctuations, e.g.
\beqa{e2} e^{-\dod t}\; \rho_\qb\, \rho_\kb\rangle = \sum_i e^{- \gd t\,
y_i \, \partial/\partial x_i}\;  e^{i (\qb+\kb)\cdot\rb_i} +
\sum_{i,j;i\ne j} e^{- \gd t\,\left(y_i \, \partial/\partial x_i + y_j
\, \partial/\partial x_j
\right)}\;  e^{i \qb\cdot\rb_i + i \kb\cdot\rb_j} \nonumber\\
= \sum_i e^{- \gd t\, y_i \, (q_x+k_x)}\;  e^{i (\qb+\kb)\cdot\rb_i}
+ \sum_{i,j;i\ne j} e^{- \gd t\,\left(y_i \, q_x + y_j \, k_x
\right)}\;  e^{i \qb\cdot\rb_i + i \kb\cdot\rb_j} = \rho_{\qb(t)}\,
\rho_{\kb(t)}\rangle . \eeqa  Under averaging with the
canonical distribution, the advection operator $\dod$ is not
self-adjoint. Thus the operator $e^{-\dod t}$ gives the advection of a
density fluctuation only at the right side of a correlator. Another
operator, $e^{\dodb t}$, gives the advection of a density fluctuation
at the left side: \beq{e3}
 \langle \rho^*_{\qb(t)} =  \langle \rho^*_\qb\; e^{\dodb t}\;,
\quad \mbox{ with }\; \dodb = \sum_i \rb_i\cdot \kap^T \cdot \left(
\bp_i + {\bf F}_i \right)\;. \eeq
(Note that $\dodb \neq\delta\Omega$; the latter was defined in \gl{b1}.)
When proving \gl{e3}, the presence of
the Boltzmann weight in the average needs to be recalled,  which
leads to the differential equation:
$$\partial_t\, \langle \rho_{\qb(t)}^* = \sum_i \langle
i \qb\cdot\kap\cdot\rb_i\, e^{-i\qb\cdot(1-\kap t)\cdot \rb_i} = -
\sum_i \int\!\!\! d\Gamma\, \rb_i\cdot\kap^T\cdot(\bp_i-{\bf F}_i)\,
\rho_{\qb(t)}^* \Psi_e = \langle \rho_{\qb(t)}^*\, \dodb\; ,$$ where
the last equality follows from partial integration. Its solution is
given by \gl{e3}. Analogously to \gl{e2}, the operator $\dodb$ also
describes the affine motion of higher order density fluctuations at the left side of an
average, so that $\langle \rho^*_{\qb(t)}\, \rho^*_{\kb(t)} = \langle
\rho^*_{\qb}\, \rho^*_{\kb}\; e^{\dodb t}$.

The transient density correlators of \gl{d1} can now be
rewritten using \gl{e1} as \beq{e4} \Phi_\qb(t) = \frac{1}{N S_q}\;
\langle \rho_{{\qb}}^* \;
 e^{\smopb t}\; e^{-\dod t} \rho_{\qb} \rangle\; .
\eeq Introducing the abbreviation $U(t)\equiv e^{\smopb t}\; e^{-\dod t}$, we find by differentiation of \gl{e4} the following result: \beq{e5}
\partial_t \; U(t)=
\partial_t \;\left( e^{\smopb t}\; e^{-\dod t} \right) =
 e^{\smopb t}\;\left(\smopb - \dod \right)\; e^{-\dod t}
 =  e^{\smopb t}\; \smoe\; e^{-\dod t}\; .
\eeq Interestingly, the equilibrium SO appears in this expression. However, because $\smoe$
and $\dod$ do not commute\rev{,} the time dependence of the correlators under shear cannot be written in terms of $\smoe$ alone (in complete accord with physical expectation). Nonetheless, the appearance here of the Hermitian equilibrium SO suggests we might
search for another `well behaved' time evolution operator to simplify the
equations of motion for the $\Phi_\qb(t)$ prior to subsequent approximation.

As will be argued further below, one candidate for a `well behaved'
time evolution operator is the following: \beq{e6} \smoa{t} \equiv
\; e^{\dodb t}\; \smoe\; e^{-\dod t}\; . \eeq  For density
fluctuations and arbitrary products of them, this possesses
identical matrix elements to the equilibrium SO, $\smoe$, with the
sole difference that all wavevectors are replaced by their
time-advected analogues (see e.g. \gl{e14} below). The SO $\smoa{t}$
in fact arises naturally in the correlator equations if one chooses
a projection onto time-dependent density fluctuations, which is the
route we now follow. (This differs from \citep{Fuchs2002c}; see
Appendix B.) This projection is achieved with the time dependent
(Hermitian) operator \beq{e7} P(t) = \sum_\qb\; \rho_{\qb(t)}
\rangle \, \frac{1}{N S_{q(t)}}\, \langle \rho^*_{\qb(t)}\; , \eeq
which clearly obeys $P(t)^2 = P(t)$.  The projector $P(t)$ can be
rewritten as \beq{e8} P(t) = \sum_\qb\; e^{-\dod t}\, \rho_{\qb}
\rangle \, \frac{1}{N S_{q(t)}}\, \langle \rho^*_{\qb}\, e^{\dodb t}
= e^{-\dod t}\, \bar P\, e^{\dodb t}\; , \eeq where the `rescaled
density projector'
$$\bar P= \sum_\qb\;  \rho_{\qb} \rangle \, \frac{1}{N S_{q(t)}}\,
\langle \rho^*_{\qb}$$  is introduced. The form of this is very
close to $P(0) \equiv P$, which is the familiar equilibrium
projector of \gl{c37}. However, $\bar P$ is normalized differently,
as a result of which it is not a true projector ($\bar P^2\ne\bar
P$).

Using
$P(t)$ and its complement  $Q(t)=1-P(t)$, \gl{e5} can be rewritten
as \beq{e9}
\partial_t \; U(t) =
 e^{\smopb t}\; \left( P(t) + Q(t) \right)\; \smoe\; e^{-\dod t}  =
\left( e^{\smopb t}\; e^{-\dod t} \right)\; \left( \bar P  \,
\smoa{t} + \smor{t} \right) = U(t) \; \left( \bar P \, \smoa{t} +
\smor{t} \right)\; , \eeq where, as promised, the SO $\smoa{t}$ now
appears. So does a further SO which represents a `remainder' term:
\beq{e10} \smor{t}= e^{\dod t}\; Q(t) \; \smoe\; e^{-\dod t}\; .
\eeq This can in turn be separated into two pieces,
$\smor{t}=\smoq{t}+\smos{t}$, one that does not couple linearly to
density fluctuations ($P\,\smoq{t}=0$) and another that does ($P
\smos{t}\ne0$). As shown in Appendix A these operators obey
\beq{e108} \smoq{t}=e^{\dodb t}\; Q(t) \; \smoe\; e^{-\dod t}\;
,\quad \smos{t}=e^{\dodb t}\; \Sigma(t)\; Q(t) \; \smoe\; e^{-\dod
t}\; , \eeq \beq{e105} \Sigma(t) =  \gd \, \int_0^tdt'\;  e^{-\dodb
t'}\, \sigma_{xy} \,  e^{\dod t'}  \; . \eeq Note that $\smoq{t}$
reduces to $Q\smoe$ in the absence of shearing, whereas $\smos{t}$
vanishes altogether in that limit.

The decomposition of $\partial_t U$ made in \gl{e9} is thus far
purely formal. However, we will find below that standard reasoning
in the framework of projection operator manipulations can be applied
to this equation \citep{Kawasaki1973b}, and that it yields, after
mode coupling approximations, a numerically stable, self-consistent
equation of motion for the transient density correlator. (Notably,
this holds even though the SO $\smor{t}$ under shear does not live
in the space perpendicular to linear density fluctuations.)

We now follow this standard reasoning, whose first step is to introduce a reduced time
evolution operator $U_r(t,t')$ that satisfies the homogeneous
differential equation for $t>t'$
$$\partial_t\; U_r(t,t') = U_r(t,t')\; \smor{t}\; ,$$
with initial value $U_r(t,t)=1$.  Its formal solution can be given
employing the time-ordered exponential function $e_-$ (see Appendix
A) where operators are ordered from left to right as time increases
\citep{Kawasaki1973b} \beq{e11} U_r(t,t') = e_-^{\int_{t'}^t ds\;
\smor{s}} = \exp_-{\left\{\int_{t'}^t ds\; \left(  e^{\dod s}\;
Q(s)\;\smoe\; e^{-\dod s} \right)\right\}}\; . \eeq Second, by
appealing to the uniqueness of the solution of the differential
equation  (\ref{e9}), and by considering the term containing $\bar
P$  as an inhomogeneity, it follows that
$$
U(t) = U_r(t,0) + \int_0^t dt'\;
U(t') \; \bar P \; \smoa{t'}\; U_r(t,t') \; .
$$
Taking a time derivative of this equation leads to the following
useful relation for the time evolution operator $U(t)$ appearing in
\gl{e4}: \beq{e12}
\partial_t \; U(t) =
U_r(t,0) \, \smor{t} + U(t)\, \bar P\, \smoa{t} + \int_0^tdt'\;
U(t')\, \bar P\, \smoa{t'}\, U_r(t,t')\, \smor{t}\; . \eeq In a third step,
the Zwanzig-Mori-type equation of motion for the transient density
correlator follows from this by taking matrix elements between density
fluctuations $\rho_\qb$.
Retaining all non-vanishing terms then gives:
\beq{e13}
 \partial_t \Phi_\qb(t) + \Gamma_\qb(t)\,
\Phi_\qb(t) +  \int_0^t dt'\; M_\qb(t,t') \;  \Phi_\qb(t') =
\Delta_\qb(t) \; . \eeq

Note that the  term on the right hand side of \gl{e12} does not
contribute in situations close to equilibrium, but remains nonzero
in the present case. This is because the projector $Q(t)$ eliminates
coupling to density only at $t=0$; it does not eliminate the
$\Sigma(t)$ contribution in \gl{e108} which subsequently acquires a
nonzero value. The resulting contribution is denoted in \gl{e13} as
\beq{e125} \Delta_\qb(t)= \frac{1}{NS_q}\; \langle \rho^*_\qb\,
U_r(t,0) \, \smor{t}\,\rho_\qb \rangle = \partial_t \langle
\rho^*_\qb\, U_r(t,0)\, \rho_\qb \rangle\; ,\eeq which indeed has
the property  $\Delta_\qb(t\to0)=\langle \rho^*_\qb\, Q\, \rho_\qb
\rangle=0$. The presence of this term stems directly from our use
here of the time-dependent projection $P(t)$, and represents (to us)
the only apparent drawback of the current approach compared to the one by
\citet{Fuchs2002c}. The advantage of the revised approach is however
revealed on examining the  instantaneous friction or `initial decay
rate', $\Gamma_\qb(t)$, in the memory equation \gl{e13}. This is
defined by the matrix element of $\smoa{t}$: \beq{e14} \Gamma_\qb(t)
=  - \frac{\langle \rho^*_\qb  \; \smoa{t} \; \rho_\qb \rangle}{N
S_{q(t)}} = - \frac{\langle \rho^*_{\qb(t)}  \; \smoe \;
\rho_{\qb(t)} \rangle}{N S_{q(t)}} =\frac{q^2(t)}{S_{q(t)}}\; .\eeq
Thus, $\Gamma_\qb(t)$ coincides with the equilibrium result 
\citep{Pus:91}, where, however, the advected wavevector replaces the
static one. This expression correctly recovers the `Taylor
dispersion' familiar for non-interacting particles (where
$M_\qb(t,t')=\Delta_\qb(t)=0$ holds). Moreover, it guarantees that
the initial decay rate in the memory equation \gl{e13} cannot become
negative, since it always corresponds to the initial equilibrium
decay rate of {\em some} wavevector. (This eliminates an important
source of numerical instability once mode coupling approximations
are applied.) Finally, the memory function in \gl{e14} is given by
\beq{e15} M_\qb(t,t') = \frac{-1}{N S_{q(t)}}  \; \langle
\rho^*_\qb\; \smoa{t'} \; U_r(t,t') \; \smor{t}\; \rho_\qb \rangle
\; , \eeq and corresponds to a generalized diffusion kernel.

We now further rearrange these exact equations by connecting the
generalized diffusion kernel to a generalized friction kernel
$m_\qb(t,t)$ which we expect to be amenable to mode coupling
approximations \citep{Fuchs2005a}. (This step is once again a
standard element of the quiescent-state MCT: since both are
non-negative, it is easier to capture a divergence in the friction
than a vanishing of the diffusivity.) The time-dependent SO is
decomposed into reducible ($\Omega_{rr}^\dagger(t)=\smor{t} \,
\tilde P(t)$) and irreducible ($\Omega_i^\dagger(t)= \smor{t}\,
\tilde Q(t)$) contributions, using the non-Hermitian projector
$\tilde P(t)$ defined by \beq{e155} \tilde P(t) = 1 - \tilde Q(t) =
\sum_{\qb}\rho_\qb \rangle\; \frac{1}{\langle \rho^*_\qb\, \smoa{t}
\, \rho_\qb \rangle } \; \langle \rho^*_\qb \; \smoa{t} \; .\eeq
This is idempotent, $\tilde P \,\tilde P = \tilde P$, and exhibits
the following couplings to density: $P\,\tilde P = \tilde P$ and
$\tilde P\, P = P$, with $P$ from \gl{c37}. (It follows that $\tilde
Q\, P = 0$.) Using it, one obtains \beq{e16} \smor{t} =
\smor{t}\left(\tilde Q(t)+\tilde P(t) \right) = \Omega_i^\dagger(t)
+ \smor{t}\, \sum_\qb \rho_\qb \rangle\; \frac{1}{\langle \rho^*_\qb
\; \smoa{t} \; \rho_\qb \rangle}\; \langle \rho^*_\qb\, \smoa{t} =
\Omega_i^\dagger(t) + \Omega_{rr}^\dagger(t)\; , \eeq  where the
irreducible part is $\Omega_i^\dagger(t) = \smor{t}\, \tilde Q(t)$
and the reducible one can be written as:
$$\Omega_{rr}^\dagger(t) = -\sum_\qb \; \smor{t}\, \rho_\qb \rangle\;
\frac{1}{N S_q \Gamma_\qb(t)}\; \langle \rho^*_\qb\, \smoa{t} \; .
$$ The decomposition of $\smor{t}$ leads to the following
differential equation for the reduced dynamics:
$$\partial_t U_r(t,t') = U_r(t,t') \; \smor{t} = U_r(t,t')\, \left( \Omega_i^\dagger(t) + \Omega_{rr}^\dagger(t) \right)\; ,
$$
which --- by the same arguments as given in support of \gl{e12}
--- has the solution \beq{e17} U_r(t,t') =  U_i(t,t') + \int_{t'}^t
dt'' \; U_r(t'',t') \, \Omega_{rr}(t'')\, U_i(t,t'') \eeq where the
irreducible \citep{Cichocki87,Kawasaki95,Fuchs2005a} fast dynamics
$U_i(t,t')$ corresponds to the solution of the corresponding
homogeneous equation. This is given  by \beq{e18} U_i(t,t') =
e_-^{\int_{t'}^t ds\; \Omega_i^\dagger(s)} =
\exp_-{\left\{\int_{t'}^t ds\; e^{\dod s}\; Q(s)\;\smoe\; e^{-\dod
s}\;
\tilde Q(s) \right\}}\; . \eeq Inserting the expression for $U_r(t,t')$
into the definition of $M_{\bf q}(t,t')$ in \gl{e15} leads to
\beq{e19} M_\qb(t,t') + \Gamma_\qb(t) \; m_\qb(t,t') \;
\Gamma_\qb(t')  + \Gamma_\qb(t) \int_{t'}^t\!\!\! dt''\;
 m_\qb(t,t'') \; M_\qb(t'',t')  = 0  \; ,
\eeq where the friction kernel is defined as: \beq{e20} m_\qb(t,t')
=    \frac{1}{N S_{q(t')}\, \Gamma_\qb(t)\, \Gamma_\qb(t')}  \;
\langle \rho^*_\qb\; \smoa{t'} \; U_i(t,t') \; \smor{t}\; \rho_\qb
\rangle \; . \eeq (Its time-dependence is given by the irreducible
dynamics introduced in \gl{e18}.)

By an entirely analogous procedure, the term $\Delta_\qb(t)$ is replaced under the irreducible dynamics by
\beq{e205} \tilde\Delta_\qb(t)= \frac{1}{NS_q}\; \langle
\rho^*_\qb\, U_i(t,0) \, \smor{t}\,\rho_\qb \rangle
 \; , \eeq which
satisfies \beq{e207} \Delta_\qb(t) = \tilde\Delta_\qb(t) -
\Gamma_\qb(t) \int_{0}^t\!\!\! dt'\;
 m_\qb(t,t') \; \Delta_\qb(t') \; .
\eeq

We are nearing the end of our task, which was to write exact equations of motion for the transient density correlators in a form suitable for mode coupling approximations. To finish it off, we now view the Zwanzig-Mori type equation of motion (\ref{e13}) as a
Volterra integral equation of second kind for $\Phi_{\bf q}(t)$,
with kernel proportional to $M_{\bf q}(t,t')$, and treat
$\left(\Delta_\qb(t)-\partial_t\, \Phi_{\bf
q}(t)\right)/\Gamma_\qb(t)$ as an inhomogeneity:
$$
\Phi_\qb(t) +  \int_0^t dt'\; \frac{1}{\Gamma_\qb(t)}M_\qb(t,t') \;  \Phi_\qb(t') =
\frac{1}{\Gamma_\qb(t)}\; \left( \Delta_\qb(t) - \partial_t \Phi_\qb(t)  \right) \; .
$$
From the theory of Volterra integral equations \citep{Tricomi57} and
the relation of $M_\qb(t,t')$ to $m_\qb(t,t')$ in \gl{e19}, one then
finds that the solution of this integral equation is obtained with
the new memory function $m_\qb(t,t')$, and is given by:
$$
\Phi_\qb(t) =  \frac{1}{\Gamma_\qb(t)}\;\left( \Delta_\qb(t) -
\partial_t \Phi_\qb(t) \right)  + \int_0^t dt'\; m_\qb(t,t') \;
\Gamma_\qb(t')\;  \frac{1}{\Gamma_\qb(t')}\; \left( \Delta_\qb(t')
-
\partial_{t'}  \Phi_\qb(t') \right)\; .
$$
This result can be brought into familiar form using \gl{e207} to
yield the final, formally exact equation of motion for the transient
density correlators: \beq{e21}
\partial_t \Phi_\qb(t) + \Gamma_\qb(t) \; \left\{
\Phi_\qb(t) + \int_0^t dt'\; m_\qb(t,t') \; \partial_{t'}\,
\Phi_\qb(t') \right\} = \tilde \Delta_\qb(t) \; . \eeq This is of
precisely the form used in quiescent-state MCT, except for the
$\tilde\Delta$ term on the right. The shear rate enters in this
term, and in the friction kernel $m_\qb(t,t')$. Suitable
approximations for these quantities are discussed in the next
Section. \rev{If $\Phi_\qb(t)$ is replaced by $\Phi_\qb^{\rm
(I)}(t)$, as defined by \citet{Fuchs2002c} and considered in
Appendix B, and if the irreducible dynamics is obtained from
time-independent projections, then the previous ITT equations of
\citet{Fuchs2005a} are recovered.}

\subsection{Mode coupling vertex}

Equation (\ref{e21}) is exact but can only be evaluated after making
approximations for the generalized friction kernel (memory kernel)
$m_\qb(t,t')$, and for the coupling of densities to fluctuating
forces, $\tilde \Delta_\qb(t)$. We seek ones that will lead to
closed equations for these quantities and for $\Phi_\qb(t)$.

We start by considering the memory kernel, which is is a correlation
function of the fluctuating forces acting on the densities, and can
be written\remrev{ more suggestively as} \rev{in a more symmetrical
fashion} \beqa{f1} m_\qb(t,t') &=& \frac{S_{q(t)}}{q^2(t)\,
q^2(t')}\, \frac 1N  \; \langle \rho^*_{\qb(t')}\; \smoe \; e^{-\dod
t'}\; U_i(t,t') \; e^{\dod t}\;
 Q(t)\; \smoe\; \rho_{\qb(t)} \rangle \nonumber\\
 &=& \frac{S_{q(t)}}{q^2(t)\, q^2(t')}\, \frac 1N  \;
\langle \rho^*_{\qb(t')}\; \smoe \, \left( 1 - \bar \Sigma(t')
\right)\, e^{-\dodb t'}\; U_i(t,t') \; e^{\dod t}\;
 Q(t)\; \smoe\; \rho_{\qb(t)} \rangle \; .
\eeqa The coupling to the stress tensor again arises for arguments
analogous to \gl{e105}, and here is given by:
$$ \bar \Sigma(t) = \gd \int_0^tdt'\; e^{\dodb t' }\; \sigma_{xy}\;e^{-\dod t' }\; ,
$$
whereas the irreducible dynamics is governed, using the
decomposition of $\smor{t}$ from \gl{e108}, by the evolution
operator: \beq{e165} \smoi{t} = \left( \smoq{t} + \smos{t} \right)
\tilde Q(t) \; .\eeq We note that $\smoq{t}\, \tilde Q(t)\, P = 0 =
P \, \smoq{t}\,\tilde Q(t)$ holds, showing that this time evolution
operator finally lives in the space perpendicular to density
fluctuations. Thus, Eqs. (\ref{e205},\ref{f1},\ref{e165}) describe
the dynamics in the space perpendicular to (linear) density
fluctuations, but also include couplings to densities via the
integrated stress tensors $\Sigma(t)$ (via the $\smos{t}$ term in
\gl{e165}) and $\bar \Sigma(t)$ in \gl{f1}.

Both $\Sigma$ terms vanish initially, but on startup of steady shear
increase linearly with accumulated strain, $\gd t$. Accordingly,
their importance depends on whether or not structural relaxation
occurs prior to the accumulation of large strains. If the
shear-induced decay of the density correlators is completed for
small values of $\dot\gamma t$, then such terms can be neglected in
a first approximation. This would be consistent with numerical
experiments by \citet{Miy:04,Var:06,Var:06b,Zau:08}, which find that
the steady state structure factor is only modestly distorted when
glasses are sheared just beyond their yield stress. (Equivalently,
the yield strain is small.) Three further arguments for the neglect
of these terms can be given. First, in the alternative approach
originally presented by \citet{Fuchs2002c}, which does not involve
time-dependent projections, these terms do not arise under shear. It
may yet be possible to establish a projection scheme in which they
vanish entirely, while maintaining the desirable positivity of the
initial decay rates $\Gamma_\qb$, captured by the present approach.
Secondly, the form of the $\tilde \Delta$ term in \gl{e21} bears
some resemblance to the `activated hopping' terms that have been
examined in several extensions of standard MCT \citep{Goe:87}, whose
goal is to replace the ideal glass transition with one in which the
relaxation time in the glass phase remains finite (but extremely
long). Arguably it is in keeping with a rheological theory of the
ideal glass transition to neglect all such terms, not just the
familiar static ones. Finally, if the yield strain is not small (so
that distortions to the structure factor remain substantial as
$\dot\gamma \to 0$), MCT's implicit use of a harmonic free energy
functional for density fluctuations is called into question; the
resulting anharmonicities cannot be captured by a theory having the
static structure factor $S_\qb$ as the only input. The $\Sigma$
terms could then consistently be neglected alongside these
anharmonicities. With these considerations in mind, from now on we
set
 \beq{f105}
\Sigma(t)\simeq 0\simeq \bar \Sigma(t)\; .\eeq However, we accept
that this approximation goes beyond those that parallel the
traditional MCT for quiescent states, and might need to be improved
upon in future work.

From this assumption, it follows as shown in Appendix A that
$U_i(t,t') = U_i^Q(t,t')$ which immediately leads to
\beq{f107}\tilde \Delta_\qb(t) \approx \frac{1}{NS_q}\; \langle
\rho^*_\qb\, U_i^Q(t,0) \; \smoq{t} \,\rho_\qb \rangle \equiv 0\;
.\eeq Moreover, on neglect of $\bar\Sigma$, only the components of
$\smoe\rho_{\qb(t)}$ perpendicular to density fluctuations enter on
either side of the time evolution operator in \gl{f1}, which means
that $m_\qb(t,t')$ can be written more symmetrically as: \beq{f2}
m_\qb(t,t') \approx \frac{S_{q(t)}}{q^2(t)\, q^2(t')}\, \frac 1N  \;
\langle \rho^*_{\qb(t')}\; \smoe \; Q(t')\;  e^{-\dodb t'}\;
U^Q_i(t,t') \; e^{\dod t}\;
 Q(t)\; \smoe\; \rho_{\qb(t)} \rangle \; .
\eeq The quantities $Q(t) \smoe\rho_{\qb(t)}$ are the fluctuating
forces. Because they do not couple to density fluctuations, in
accord with our mode coupling precepts, we approximate them by their
overlap with pairs of densities. The appropriate projection, which
is time dependent due to wavevector advection in \gl{f2}, is then
\beq{f3} P_2(t) = \sum_{\kb>\pb,\kb'>\pb'} \frac{\rho^{}_{\kb(t)}
\rho^{}_{\pb(t)} \rangle \langle \rho^*_{\kb'(t)} \rho^{*}_{\pb'(t)}
}{ \langle \rho^*_{\kb(t)} \rho^*_{\pb(t)} \rho^{}_{\kb'(t)}
\rho^{}_{\pb'(t)} \rangle} \approx \sum_{\kb>\pb}
\frac{\rho^{}_{\kb(t)} \rho^{}_{\pb(t)} \rangle \langle
\rho^*_{\kb(t)} \rho^{*}_{\pb(t)} }{ N^2 S_{k(t)} S_{p(t)}} \; ,
\eeq where a Gaussian decoupling of the four-point density
fluctuation gives the second form, and the wavevector inequalities
prevent overcounting.

Using this in \gl{f2} for the fluctuating forces gives
\beq{f4}
m_\qb(t,t') \approx   \frac{S_{q(t)}}{q^2(t)\, q^2(t')}\, \frac 1N  \;
\langle \rho^*_{\qb(t')}\; \smoe \; Q(t')\; P_2(t')\;  e^{-\dodb t'}\;
 U^Q_i(t,t') \; e^{\dod t}\; P_2(t)\;
 Q(t)\; \smoe\; \rho_{\qb(t)} \rangle
\eeq Finally, we factorise the four-point correlation function with
reduced dynamics into the product of correlators with full dynamics,
just as was discussed in connection with \gl{d36}: \beqa{f5} \langle
\rho^*_{\kb(t')}\;\rho^*_{\pb(t')}\;   e^{-\dodb t'}\; U^Q_i(t,t')
\; e^{\dod t}\; \rho^{}_{\kb'(t)}\; \rho^{}_{\pb'(t)} \rangle
\approx  N^2\, S_{k(t')}\, S_{p(t')}\; \Phi_{\kb(t')}(t-t')\;
\Phi_{\pb(t')}(t-t')\; \delta_{\kb,\kb'}\,\delta_{\pb,\pb'}\; .
\eeqa The Kronecker-$\delta$'s arise from translational invariance
and homogeneity.  The vertex functions $V_{\qb\kb\pb}$ measure the
overlap of the fluctuating forces with the density pair fluctuations
in \gl{f4} and can be evaluated from equilibrium information.
Although they are required as functions of the advected wavevectors,
their evaluation proceeds exactly as in equilibrium MCT in
\citet{Goe:91} \beq{f6} V_{\qb\kb\pb} = \frac{\langle \rho^*_{\qb}\;
\smoe\; Q\; \rho_{\kb}\; \rho_{\pb} \rangle}{N S_{k} S_{p}} =
\qb\cdot \left( \kb \, n c_k + \pb\, n c_p \right)\;
\delta_{\qb,\kb+\pb} \; , \eeq where $c_q$ is the equilibrium direct
correlation function connected to the structure factor via the
Ornstein-Zernike equation $S_q=1/(1-nc_q)$. (As is standard practice
 the convolution approximation has been used in
\gl{f6} to neglect a small contribution from higher order direct
correlations \citep{Goe:91}.)

The final expression for the memory
function in our mode coupling approximation can then be written:
\beq{f7} m_\qb(t,t') =  \frac{1}{2N} \sum_{\kb} \frac{S_{q(t)}\,
S_{k(t')}\, S_{p(t')}}{q^2(t)\; q^2(t')}\;
 V_{\qb\kb\pb}(t)\, V_{\qb\kb\pb}(t')\;
 \Phi_{\kb(t')}(t-t')\;  \Phi_{\pb(t')}(t-t')\; .
\eeq Here  $\pb\equiv\qb-\kb$, and we have abbreviated
$V_{\qb\kb\pb}(t)=V_{\qb(t)\kb(t)\pb(t)}$. A change of the
integration-variable from $\kb$ to $\kb'=\kb(t')$ leads to an
alternative expression which \mike {may exhibit more clearly} the various origins of the
time-dependences. We write: \beq{f7b} m_\qb(t,t') =
\bar m_{\qb(t')}(t-t')\; , \eeq where the reduced memory function
$\bar m_{\qb'}(\tau)$ is evaluated at the time-dependent wavevector
$\qb'=\qb(t')$, and depends on time only via the difference
$\tau=t-t'$. This quantity is given by: \beq{f8} \bar m_{\qb'}(\tau)
= \frac{1}{2N} \sum_{\kb'} \frac{S_{q'(\tau)}\, S_{k'}\,
S_{p'}}{{q'}^2(\tau)\; {q'}^{2}}\;
 V_{\qb'\kb'\pb'}(\tau)\, V_{\qb'\kb'\pb'}\;
 \Phi_{\kb'}(\tau)\;  \Phi_{\pb'}(\tau)\; .
\eeq Here, $\pb'=\qb'-\kb'$ holds, with $\qb(t)=\qb'(\tau)$ and
analogous expressions for the other wavevectors.  \disc{[THERE IS NO
$\kb$ OR $\pb$ IN THE EQUATION SO NO EXPRESSIONS SHOULD BE GIVEN FOR
THEM. ANYWAY, THE CURRENT NOTATION SIMPLY DOES NOT WORK. IT REQUIRES
SIMULTANEOUSLY $\qb'=\qb(t')$ AND $\qb(t)=\qb'(\tau)$. WHY ARE THERE
PRIMES IN \gl{f8}? NOTHING IS SHOWN `MORE CLEARLY' BY THIS EQUATION
IN ITS CURRENT FORM.] It is supposed to mean: $\qb'=\qb(t')$ defines
the vector $\qb'$. It gets advected for a time $\tau$ to
$\qb'(\tau)$. This turns out to be $\qb'(\tau)=\qb'(1-\kappa
\tau)=\qb(1-\kappa t')(1-\kappa(t-t'))=\qb(1-\kappa t)=\qb(t)$. How
should I clarify?}

Notably, the mode coupling vertex derived from field theory
\citep{Miy:02,Miy:04} for the time-dependent fluctuations around the
stationary state \mike{apparently coincides}\disc{ [WHAT DOES THAT
MEAN?] their presentation is even less clear than mine ;-), and
their result possibly agrees with us; yet, it depends on the
interpretation of their formulae.} with \gl{f8}, albeit with the
important difference that the distorted structure factor enters
there in place of our static one. (The different philosophies behind
either approach were discussed in the Introduction.) If the
`external' wavevector $\qb$ lies in the plane perpendicular to the
flow direction, $\qb \cdot \hat{\bf x} = 0$, and thus is not
advected, $\qb(t')=\qb'=\qb$, then the memory function simplifies
further\rem{. While the `internal' wavevector $\pb'=\pb(t')$ in
general is time-dependent, in this case it simplifies to a
time-independent wavevector ${\pb'}=\qb-{\kb'}$, and the remaining
dependence on time enters $m_\qb(t,t')$}\add{, as time enters} only
via the time difference $\tau=t-t'$, $m_{\qb; q_x = 0}(t,t')=\bar
m_{q_y,q_z}(\tau)$ . \disc{[I HAVE GIVEN UP TRYING TO UNDERSTAND
THIS PARAGRAPH EVEN THOUGH I KNOW WHAT IT IS TRYING TO SAY. ADVISE
REWRITE AFTER THE ABOVE NOTATION IS MADE CORRECT.] please take a
look at the changed text.}

The generalized friction kernel of \gl{f7}, which vanishes in the
absence of particle interactions, scales like $q^2$ for small
wavevectors. Whereas one power in $q$ arises from density
conservation, the second arises because the total force among all
particles vanishes via Newton's laws. (The present result for
$m_\qb(t,t')$ of course recovers the standard MCT expression for
$\gd=0$, to which similar remarks apply.) Viewed as a function of
$t$, the memory function under shear has a positive maximum value at
$t=t'$, which describe instantaneous friction. As $t'$ is increased,
this value decreases under the cumulative effects of shearing
between startup and time $t'$. Shear decreases the correlations by a
dephasing of the two vertex factors which enter \gl{f7}. The two
factors coincide only at $t=t'$, creating a squared vertex familiar
from standard MCT. (At other times, the product is not necessarily
positive so that stability of \gls{e21}{f7} is not automatic.) This
dephasing results from a shift of the (internal) advected
wavevectors to higher values, thus suppressing the effective
interaction potentials $c_{k\to\infty}\to0$ and decreasing the
friction. Additional decorrelation during the time interval $t-t'$
enters via the density correlators; this represents a `Brownian
decay factor' (in the sense that these correlators do not decay
without Brownian motion; see Section \ref{tradens}).  The overall
effect of shearing, as previously discussed elsewhere
\citep{Fuchs2002c,Fuchs2003} is to cut off memory and thereby
fluidize the system.

\section{Summary and discussion }
\label{disc}

Our combined ITT/MCT approach to the rheology of steadily sheared
suspensions consists of the approximated generalized Green-Kubo
relations summarized in Section \ref{green}, which introduce the
transient density correlator (defined in \gl{d1}) describing
structural relaxation, and its  equation of motion, which results
from \gl{e21}: \beq{g1}
\partial_t \Phi_\qb(t) + \Gamma_\qb(t) \; \left\{
\Phi_\qb(t) + \int_0^t dt'\; \bar m_{\qb(t')}(t-t') \;
\partial_{t'}\,  \Phi_\qb(t') \right\} = 0 \; . \eeq
The initial decay rate is given in \gl{e14} and the mode
coupling approximation for the memory function in \gl{f8}.

Since the preliminary presentation of the ITT approach by
\citet{Fuchs2002c}, a number of results found using various
formulations and simplifications, based on this general framework,
have been worked out \citep{Fuchs2003,Haj:08}. All these are
underpinned by the more complete presentation offered here, which
however differs from the original version, as mentioned previously
and summarized in Appendix B. These differences do not affect the
main conclusions from the ITT approach, and it lies beyond our scope
to present details of all of \mike{such predictions} here. Below, we
summarize some of the most important ones, point to the literature
for more detailed discussions, and compare some additional
predictions with recent experiments and simulations. Because the
transient density correlators $\Phi_\qb(t)$ determine the steady
state properties, we start with them. Let us recall, from \gl{d17},
that within the current mode coupling scheme transient and steady
state density correlators essentially coincide, being connected via
$C_\qb(t,\gd)=S_\qb(\gd)\, \Phi_\qb(t)$. Although as mentioned in
Section \ref{coupling}.3 this is only an approximation, \rev{using} it allows
comparison with a wider range of experimental and simulation data.

\subsection{Universality aspects}

The ITT/MCT equations require as input the equilibrium structure
factor $S_q$, the average density $n$, and the shear rate $\gd$. The
dynamics on short time scales, which is not treated adequately in
ITT,  is specified only by the bare (short time) diffusion
coefficient $D_0$ which enters $\Gamma_\qb(t)$ as a factor (we set
$D_0 = 1$). The ITT equations contain a bifurcation
 in the long time limits $f_\qb$ of the
transient density correlators, $\Phi_\qb(t\to\infty)=f_\qb$. This
nonequilibrium transition generalizes the ideal glass transition
found in standard MCT, and its predicted position as a function of
density, temperature etc (entering via $S_q$) coincide with the
standard one: there is no shift caused by $\gd$. Nonetheless,  all
nonergodic states become ergodic under infinitesimal shearing, that
is, the glass `form factors' vanish, $f_\qb\equiv 0$ for $\gd\ne 0$.
Thus, enforcing even an infinitesimal steady shear flow melts the
glass, creating a state in which all density fluctuations --
including those not directly advected by shear -- relax in finite
time. This follows simply because shear advection cuts off the
structural memory and forces $\bar m_\qb(t\to\infty)=0$.

A true nonergodicity transition, like the ideal glass transition of
standard MCT \citep{Goe:91,Goe:92}, thus does not exist in ITT/MCT
(unless one sets $\dot\gamma = 0$ from the outset, in which one
recovers the standard theory). However, at the locus of the MCT
glass transition, ITT/MCT predicts a transition from a viscoelastic
fluid to a yielding glass. In fluid states, steady state averages,
correlators, etc. show a linear response regime in shear rate
amounting to a regular Taylor series in $\gd$\add{ (at least in low
orders)}; here, the theory reduces to MCT for $\gd\to0$. In glass
states, however, averages obtained from the generalized Green-Kubo
relations do not possess a linear response regime. This follows
because the final relaxation of the transient correlators now arises
{\em only} because of the drive by shearing. The integrals in the
Green-Kubo relations thereby attain non-analytic dependences on
$\gd$. (See Section \ref{struc} for the example of the distorted
structure factor and \citep{Fuchs2002c,Fuchs2003,Haj:08} for
the shear stress.) {\em A posteriori}, this result is not
surprising, because, as Maxwell explained (and MCT recovers) glasses
are solids and should thus exhibit a linear response regime to
strain (or deformation gradient) but not to shear rate (or velocity
gradient).

Close to the glass transition, the ITT/MCT correlators exhibit a
two-step relaxation, comprising a decay onto $f_\qb$, followed by
the relaxation to zero. The dynamics close to the plateau $f_\qb$
factorises into time-independent amplitudes and a time-dependent
scaling function \citep{Fuchs2003}: \beq{g2} \Phi_\qb(t) = f^c_q +
h_q\; {\cal G}(t)\; , \eeq where the critical glass form factor
$f^c_q$ and the so-called critical amplitude $h_q$ are isotropic,
not affected by shear, follow from the equilibrium structure factor
at the transition point, and agree with the corresponding quantities
from quiescent MCT \citep{Goe:91,Goe:92}.

The scaling function ${\cal G}(t)$ carries all the important
dependence on time, shear rate and other control parameters. In
quiescent fluid states, the final relaxation, called the
$\alpha$-process of glassy relaxation, is isotropic, and possesses a
finite internal relaxation time, to be denoted as $\tau$. This
depends on a `separation parameter' $\epsilon$ which is any
combination of thermodynamic control parameters that vanishes
linearly at the arrest transition; for colloids we choose $\epsilon
= (\phi-\phi_c)/\phi_c$ where $\phi$ denotes volume fraction and
$\phi_c$ the critical value of this. The scaling function is
initiated by a power law, called the von Schweidler law \mbox{${\cal
G}(t\to\infty,\epsilon<0,\gd=0) \propto - (t/\tau)^b$}, with
exponent $b$ smaller than unity. ITT identifies the `dressed Peclet'
or Weissenberg number Pe $=\gd\tau$ as the expansion parameter that
controls when the effect of shearing the fluid starts to matter, at
Pe $\gtrapprox \,1$ . This of course depends sensitively on
$\epsilon$ because the $\alpha$-relaxation time does so.

In states which are nonergodic glasses at rest, the relaxation time
$\tau$ formally is infinite. As already made clear, such states are
melted by any nonzero shear rate $\gd$, as can be deduced from the
nonlinear stability equation for ${\cal G}(t)$ : \beq{g3} \tilde
\epsilon - c^{(\gd)}\; (\gd t)^2 + \lambda\; {\cal G}^2(t) =
\frac{d}{dt} \int_0^t\!\!\!dt'\; {\cal G}(t-t')\; {\cal G}(t')\; ,
\eeq which, along with its initial condition \beq{g4} {\cal G}(t \to
0) \to (t/t_0)^{-a} \; , \eeq was derived via ITT by
\citet{Fuchs2002c}. The so-called critical law in \gl{g4} is the
asymptotic solution for long times right at the bifurcation point
$\varepsilon=0=\gd$, which can be matched onto the short-time
dynamics by chosing the time scale $t_0$. The parameters $\lambda$
and $c^{(\gd)}$ in \gl{g3} are determined by the static structure
factor at the transition point, and, in the case of hard-sphere
colloids, take values around $\lambda\approx 0.73$ and
$c^{(\gd)}\approx 0.65$ if one estimates $S_q$ from the
Percus-Yevick approximation  \citep{Fuchs2003}. The transition point
then lies at packing fraction $\phi_c=\frac{4\pi}{3} n_c
R_H^3\approx 0.52$, and the scaled separation parameter entering
\gl{g3} obeys $\tilde\epsilon=C\, \epsilon$ with $C \approx 1.3$.
The exponent $a$ in \gl{g4} depends on the `exponent parameter'
$\lambda$ via $\lambda=\Gamma(1-a)^2/\Gamma(1-2a)$ where $\Gamma$
denotes the Gamma function \citep{Goe:91,Goe:92}.

The presence of a finite shear rate enforces ergodic decay of all
density fluctuations, not just those directly advected by shearing.
Indeed, the long-time solution of \gl{g3} for all separation
parameters exhibits a finite-time relaxation (or `yield process')
whose initial decay from the plateau is governed by the asymptote
\citep{Fuchs2002c}: \beq{g5} {\cal G}(t \to \infty  ) \to -
\sqrt{\frac{c^{(\gd)}}{\lambda-\frac 12}} \; |\gd t| = - t /
\tau_{\gd} \; . \eeq The effect of hydrodynamic interactions (HI) on
the local dynamics, were this to be incorporated into our approach,
could be expected to shift the matching time $t_0$ in \gl{g4}. (Note
that $\gd t_0 \ll 1$ is required for the asymptotic solution
provided by the factorisation law in \gl{g2} to become valid.)
However our results for the the shear-induced relaxation time
$\tau_{\gd}\sim 1/\gd$ in \gl{g5} cannot be changed without
wholesale alterations to the structure of the theory, in which shear
only enters via the affine advection of fluctuation patterns.

Schematic models, in which all or part of the wave-vector dependence
in the theory is suppressed \citep{Fuchs2003,Haj:08} can capture all
aspects of ${\cal G}(t)$ and therefore provide an important toolbox
for exploring generic aspects of the transient dynamics in ITT. (For
the same reason, such models played an important role in the
development of standard MCT for quiescent glasses
\citep{Goe:84,Goe:91}.) In particular the F$_{12}^{(\gd)}$ model
\citep{Fuchs2003}, in which one considers only a single
representative wavevector, has been used extensively to analyse the
steady state flow curves $\sigma(\gd)$ obtained in colloidal
dispersions of hard and soft spheres and in computer simulations of
binary supercooled mixtures \citep{Fuchs2003,Cra:06,Cra:08,Haj:08}.
With rather few adjustable parameters, very good global agreement
between measured steady state data and fitted curves from the
F$_{12}^{(\gd)}$ model could be obtained. The semi-schematic
isotropically sheared hard sphere model (ISHSM) was developed to
wholly incorporate the full isotropic dynamics of quiescent MCT,
while simplifying the rather complicated shear advection effects
presented above for the ITT/MCT theory. This was done by selective
use of an `isotropic average' in which all wavevectors are
effectively treated as though pointing in the vorticity direction
\citep{Fuchs2002c,Fuchs2003}. An ISHSM model, whose results are
compared to spatially resolved data from experiments and simulations
in Section \ref{yield}, is summarized for completeness in Appendix
C.

\subsection{Yielding process}
\label{yield}

 The shear induced decay of density fluctuations causes
the transient correlator $\Phi_\qb(t)$ to fall below the plateau
$\Phi_\qb\simeq f_q$ that would persist indefinitely for an
unsheared glass. Since long time Brownian motion is frozen out, we
might expect this decay to be purely strain-induced and thus rate
independent, in which case a scaling law, $\Phi_\qb(t,\gd)\to\tilde
\Phi_\qb(\tilde t)$, should describe the yielding process. Here, the
rescaled time is $\tilde t = t / \tau_{\gd} \propto t |\gd|$. \rev{The existence of such a scaling law has been called {\it time-shear-superposition principle}  \citep{Bes:07}.} At the
transition, $\epsilon=0$, and neglecting \add{for small accumulated
strains $\gd t\ll1 $} \disc{[ON WHAT GROUNDS?], presently for lack
of proofs beyond this. it is at least consistent with the other
approximations and works as long as the decay is finished at strain
0.1 or so. It holds in the ISHSM. } the external advection in the
memory function, $\bar m_{\qb(t')}(t-t')\approx \bar m_{\qb}(t-t')$,
the equation for the master function $\tilde \Phi_{\qb}$ was found
as \citep{Fuchs2003} \beq{g6} \tilde \Phi_\qb(t) = \tilde
m_{\qb}(\tilde t) - \frac{d}{d\tilde t}\;\int_0^{\tilde t} dt'\;
 \tilde m_{\qb}(\tilde t-t') \;
  \tilde \Phi_\qb(\tilde t')  \; . \eeq
The memory function $\tilde m_{\qb}(\tilde t-t')$ is the one defined
from \gl{f8}, evaluated at $\epsilon=0$ and with the asymptotic
scaling form $\tilde \Phi_{\qb}$ replacing the actual correlator.
The result \gl{g6} highlights the accelerated decay of correlations
induced by shear. The structural decay depends on the accumulated
strain $\tilde t\propto |t\gd|$.

The yielding process is initiated by \gls{g2}{g5}, $\tilde
\Phi_\qb(\tilde t\to0,\epsilon=0)=f^c_q-h_q \tilde t$, which
describes an initially isotropic yielding process under shear. The
linear initial decay of $\tilde \Phi_q(\tilde t)$ also suggests
little stretching of the final decay. \add{This suggests an
isotropic exponential as approximation to the yielding master
function $\tilde \Phi_\qb(\tilde t')\approx
f^c_q\,\exp{\{-t/\tau(\gd,q)\}}$, with $\tau(\gd,q)=(f^c_q/h_q) \,
\tau_{\gd}$. }However, the true degree of anisotropy of the process
and its non-exponentiality at later rescaled times is still unknown.

Note that all quantities characterising the short-time motion, like
the matching time $t_0$, have dropped out of \gl{g6}. Its solutions
$\tilde \Phi_\qb(\tilde t)$, and the stationary averages obtained
from them, like the yield stress $\sigma^+$, thus in ITT depend on
the equilibrium structure factor only.\add{ \mike{Hydrodynamic corrections to the bare diffusivity}, for example, are predicted not to affect the value of
the yield stress, which at the transition can be obtained explicitly
from  $$ \sigma^+_c =
\frac{k_BT}{2}\sqrt{\frac{2\lambda-1}{2c^{(\gd)}}}
\int_0^\infty\!\!\!\!d\tilde t\, \int\!\!\frac{d^3k}{(2\pi)^3}\;
\frac{k_x^2k_yk_y(-\tilde t)}{k\, k(-\tilde t)}\;
\frac{S'_kS'_{k(-\tilde t)}}{S^2_{k}}\; \tilde \Phi^2_{\kb(-\tilde
t)}(\tilde t)\; . $$ }
 \disc{[I CANNOT REVISE THIS PARAGRAPH SINCE I HAVE NO
IDEA WHAT IT IS TRYING TO SAY.] please take a look at the added
text.}

\begin{figure}[t]
\centering
\includegraphics[ width=0.6\columnwidth]{besselingmaster.eps}
\caption{Steady state incoherent intermediate scattering functions
$\Phi^s_q(t)$ as functions of accumulated strain $\gd t$ for various
shear rates $\gd$; the data were obtained by \citet{Bes:07} in a
colloidal hard sphere dispersion at packing fraction $\phi=0.62$ (at
$\epsilon\approx 0.07$) using confocal microscopy; the wavevector
points in the vorticity ($\bf \hat z$) direction and has $q=3.8/R$
(at the peak of $S_q$).  The effective Peclet numbers Pe$_{\rm
eff}=4 R^2 \gd/D_s$ are estimated with the short time self diffusion
coefficient $D_s\approx D_0/10$ at this concentration from
\citet{Meg:98}. ISHSM calculations with separation parameter
$\epsilon=0.066$ at $qR=3.9$ (PY-$S_q$ peaking at $qR=3.5$), and for
strain parameter $\gamma_c=0.033$, are compared to the data for the
Pe$_{\rm eff}$ values labeled. The yielding master function at
Pe$_{\rm eff}=0$ lies \mike{among the experimental} data curves
which span $0.055 \le$ Pe$_{\rm eff} \le 0.45$, but discussion of
the apparent systematic trend of the experimental data would require
ISHSM to better approximate the shape of the final relaxation
process. \label{Fig1}}
\end{figure}

\begin{figure}[t]
\centering
\includegraphics[ width=0.6\columnwidth]{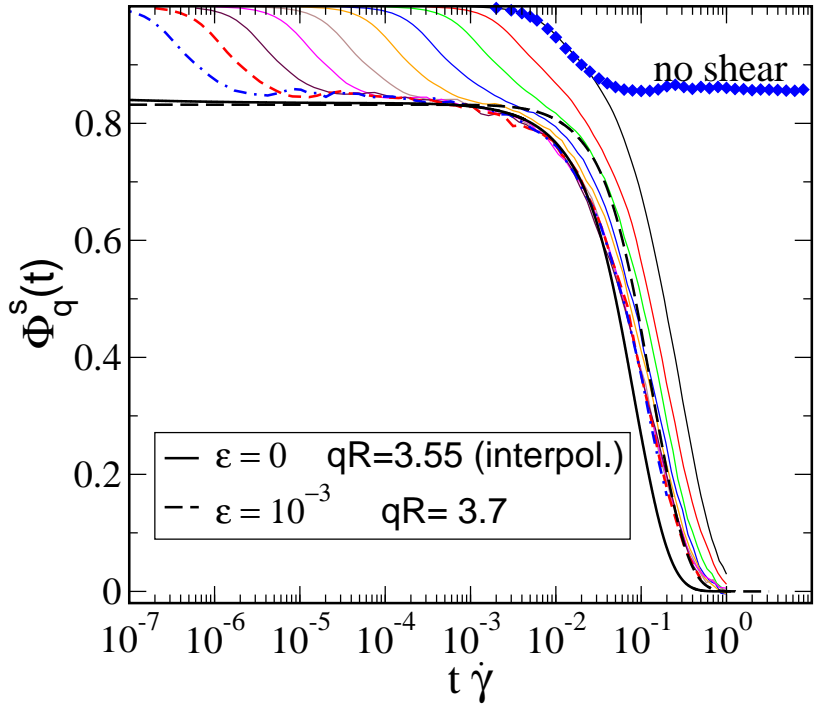}
\caption{Steady state incoherent intermediate scattering functions
$\Phi^s_q(t)$ measured in the vorticity direction as functions of
accumulated strain $\gd t$ for various shear rates $\gd$; data from
molecular dynamics simulations of a supercooled binary Lenard-Jones
mixture below the glass transition at $(T_c-T)/T_c\approx 0.3$
\citep{Var:06}. These collapse onto a yield scaling function at long
times. The wavevector is $q=3.55/R$ (at the peak of $S_q$).  The
quiescent curve, shifted to agree with the one at the highest $\gd$,
shows ageing dynamics at longer times outside the plotted window.
The apparent yielding master function from simulation is compared to
the ones calculated in ISHSM for glassy states at or close to the
transition (separation parameters $\epsilon$ as labeled) and at
nearby wave vectors (as labeled). ISHSM curves were chosen to match
the plateau value $f_q$, while strain parameters $\gamma_c=0.083$ at
$\epsilon=0$ (solid line) and $\gamma_c=0.116$ at $\epsilon=10^{-3}$
(dashed line) were used. \label{Fig2}}
\end{figure}

While data for the transient density correlators remain scarce
\citep{Zau:08}, detailed measurements of the stationary dynamics
under shear of a colloidal hard sphere glass have recently been
performed by  \citet{Bes:07} using confocal microscopy. Single
particle motion was investigated. We can therefore compare theory to
experiment for the density relaxation in a shear-melted glass  at
roughly the wavevector inverse to the average particle separation,
if we make three further assumptions: $(i)$ that transient and
stationary correlation functions agree up to an amplitude factor
(\gl{d17}); $(ii)$ that it is valid to compare the measured
incoherent density correlators to the coherent ones calculated in
ITT; and $(iii)$ that the ISHSM (Appendix C) is adequate for such a
comparison. Figure \ref{Fig1} then shows self-intermediate
scattering functions measured for wavevectors along the vorticity
direction where neither affine particle motion nor wavevector
advection appears. The stationary correlators deep in the glass, for
shear rates spanning almost two decades, are shown as function of
accumulated strain $\gd t$, to test whether a simple scaling
$\tau_{\gd}\sim 1/\gd$ as predicted by \gl{g6} holds. Small but
systematic deviations are apparent which have been interpreted as a
power law $\tau_{\gd}\sim \gd^{-0.8}$ \citep{Bes:07,Sal:08}. (ISHSM
computations were performed for a nearby wavevector where $S_q$ is
around unity so that coherent and incoherent correlators may be
assumed to be similar as argued by \citet{Pus:78}.) A separation
parameter $\epsilon$ very close to that of the experiments was taken
for the fit, allowing for a small deviation so as to match better in
amplitude the final relaxation step. The yielding master function
from ISHSM can be brought into register with the data measured at
small effective Peclet numbers Pe$_{\rm eff}$, by using a
phenomenological `strain rescaling parameter' $\gamma_c=0.033$ (see
Appendix C for the definition of this). For this parameter one
expects values of order unity to be able to compensate \rev{for} the
oversimplified treatment of angle averaging in ISHSM; the smallness
of the fitted value is not yet understood.  The effective Peclet
number Pe$_{\rm eff}=4R^2 \gd / D_s$ introduced for these fits
measures the importance of shear relative to the Brownian diffusion
time obtained from the short time self diffusion coefficient $D_s$
at the relevant volume fraction. (A value $D_s/D_0=0.1$ was taken
from \citet{Meg:98}.) The shape of the final relaxation process
differs subtly between experiment and theory, and therefore we can
make no definitive comment at this stage on the observed small
systematic drift of the rescaled experimental curves. (This drift is
responsible for the fitted dependence $\tau_{\gd}\sim \gd^{-0.8}$
reported by \citet{Bes:07}.) At the larger effective Peclet numbers,
Pe$_{\rm eff}\ge 0.5$, for which the short-time and final
(shear-induced) relaxation processes move closer together, the model
gives quite a good account of the $\gd$-dependence. Overall, theory
and experiment agree in finding a two step relaxation process, where
shear has a strong effect on the final structural relaxation, while
the short time diffusion is not much affected.

The shape of the final relaxation step in a shear-melted glass can
be studied even more closely in recent computer simulations by
\citet{Var:06}, where a larger separation of short and long time
dynamics could be achieved. In these molecular dynamics simulations
of an undercooled binary Lenard-Jones mixture, schematic ITT models
give a good account of the steady state flow curves, $\sigma(\gd)$
\citep{Var:06b}. Figure \ref{Fig2} shows the corresponding
stationary self intermediate scattering functions for a wavevector
near the peak in $S_q$, oriented along the vorticity direction,
\rev{at} shear rates spanning more than four decades. Collapse onto a
master function when plotted as function of accumulated strain is
nicely observed as predicted by \gl{g6}. At larger shear rates, the
correlators peel away from the master function; this resembles the
behaviour observed in the confocal experiments in Fig.~\ref{Fig1}.
Using the additional assumptions $(i)$ to $(iii)$ \rev{given} above, the
shape of the master function can be fitted using our ISHSM
calculations. As before, to bring these into register, a strain
parameter $\gamma_c$ was introduced, whose smallness again remains
unaccounted for. After this rescaling, modest but visible
differences in the shapes remain: the theoretical master function
decays more steeply than the one from simulations.

\subsection{Distorted structure under shear}
\label{struc}

 From the transient correlators discussed (and,
partially, validated) in the previous Section, all stationary
averages then follow within ITT-MCT approach. While the resulting
flow curves and yield stresses have been extensively predicted via
both ISHSM and schematic ITT models \citep{Fuchs2003}, the distorted
microstructure has so far only been calculated within ITT to lowest
order in shear rate \citep{Henrich2007}.
\begin{figure}[t]
\centering
\includegraphics[ width=0.6\columnwidth]{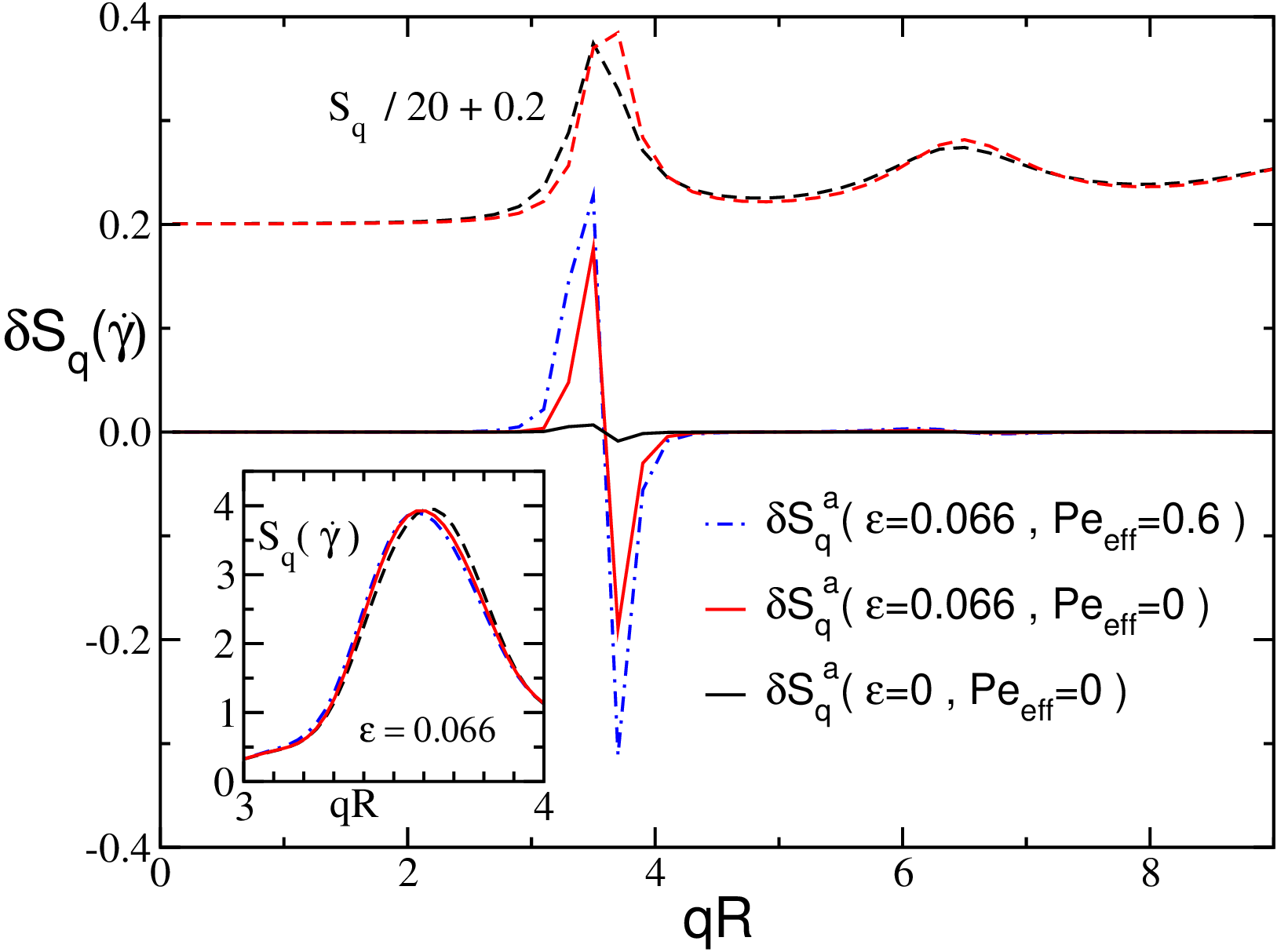}
\caption{(Isotropic) distortion of the structure factor $\delta
S_{q}(\gd)$ under shear within ISHSM approximation in glass states.
The limiting values for $\gd\to0$ (Pe$_{\rm eff}=0$) are shown at
the transition ($\epsilon=0$, black line), and at $\epsilon=0.066$
(red line) corresponding to the confocal data in Fig.~\ref{Fig1}  at
$\phi=0.62$; for the latter $\epsilon$, the increase of  $\delta
S_{\qb}(\gd)$ with shear rate is shown (blue dot-dashed line);
strain parameter $\gamma_c=0.033$ was chosen. The upper thin dashed
lines indicate the equilibrium structure factors $S_q$ from PY
approximation for these densities (with corresponding colour code).
The inset shows the complete distorted structure factors
$S_{\qb}(\gd)$ at $\phi=0.62$ (red solid line for $\gd\to0$; blue
dashed-dotted line at Pe$_{\rm eff}=0.6$) compared to the
equilibrium ones (black dashed line); Spline interpolation through
the data on the grid of the main figure was used. At $\epsilon=0$
the difference for $\gd\ne0$ can not be resolved. \label{Fig3}}
\end{figure}

As discussed in Section \ref{intro},  the microstructure possesses a
linear response to $\gd$ only in fluid states, while in yielding
glass states, the limit $S_{\qb}(\gd\to0)$ does not agree with the
equilibrium structure factor denoted $S_q$. Rather ITT-MCT provides
the general results
$$S_{\bf q}(\dot{\gamma}) = S_q + \delta S^{\rm aniso}_{\bf q}(\dot{\gamma}) +
\delta S^{\rm iso}_{q}(\dot{\gamma})\;
$$
with \mikecom{Please confirm the $S_q-1/n$ is not a misprint (here
and in appendix). Please check $c_k$ was defined previously (I think
not)} \beq{g7} \delta S^{\rm aniso}_{\bf q}(\dot{\gamma}) =
\int_0^\infty\!\!\!\!dt\; \left(\frac{d}{d t} S_{q(-t)}\right) \;
\Phi^2_{\bf q(-t)}(t)
\eeq where the substitution $ \frac{d}{d t} S_{q(-t)} = \gd\,
\frac{q_x q_y(-t)}{q(-t)}\; S'_{q(-t)} $ was made in order to tidy
up the expressions given in \gl{d13}.\add{ As discussed in
Sect.~\ref{avstruc}, we refrain from discussing $\delta S^{\rm
iso}_{q}(\dot{\gamma})$, which is quantitatively smaller for not too
large wavevectors.} The rewriting in \gl{g7} highlights how the
distorted structure arises from the affine stretching of
(equilibrium-amplitude) density fluctuations, in competition with
structural rearrangements as encoded in the transient correlators.
Equation (\ref{g7}) also shows that whenever the decay is
shear-induced (so that the final decay of $\Phi_{\qb}(t)$ depends on
time via $t/\tau_{\gd}\sim t|\gd|$) the limit $\delta
S_{\qb}(\gd\to0)\ne 0$ follows. This limit can be studied by
replacing $\Phi_{\qb}(t)$ in \gl{g7} with $\tilde \Phi_{\qb}(\tilde
t)$ from \gl{g6}, and by performing the time integrals over the
rescaled time $\tilde t\propto t|\gd|$. For the anisotropic
contribution for example one then finds the `yield value' of the
structural distortion \beq{g8} \delta S^{\rm aniso, + }_{\bf
q}(\varepsilon\ge0)\equiv \delta S^{\rm aniso}_{\bf
q}(\dot{\gamma}\to 0,\varepsilon\ge0)= \int_0^\infty\!\!\!\!d\tilde
t\; \left(\frac{d}{d\tilde t} S_{q(-\tilde t)}\right) \; \tilde
\Phi^2_{\bf q(-\tilde t)}(\tilde t)\; , \eeq where the right hand
side is explicitly shear-rate independent. The discontinuity
represented by  $\delta S_{\qb}^+(\varepsilon\ge0) \ne 0$ reflects
the same physics as the dynamic yield stress; we therefore call it
the `dynamic yield contribution' to the structure factor. Figure
\ref{Fig3} shows distorted microstructures under shear for glass
states at infinitesimal and finite shear rates. These results were
obtained within the ISHSM of Appendix C, at parameter settings
corresponding to those fit to the confocal microscopy data in
Fig.~\ref{Fig1}, so as to assess the magnitude of the effects in a
realistic case. Thanks to the isotropisation within ISHSM, the
distorted microstructure\add{ becomes isotropic in the yielding
glass}.\disc{ [CAUTION: THIS SHOULD VANISH IF BAXTER TERM IS
REMOVED!], no, also the term called 'anisotropic' becomes isotropic
for $\gd\tau\gg1$.} Right at the transition, $\epsilon=0$, only a
tiny discontinuity $\delta S_{\qb}^+$ appears. This suggests that
the dynamic yield contribution to the structure factor might not be
discernible until one moves deeper into the glass.\rem{ the dynamic
yielding contribution grows appreciably. On increasing the shear
rate the effect should also become clearly noticeable; this holds
even for the isotropic contribution [CAUTION] if the equilibrium
structure factor is known from some independent measurement.}\disc{
[DOES ISHSM ISOTROPIC CONTRIBUTION STEM FROM THE BAXTER TERM OR IS
IT SOMEHOW AN APPROXIMATION TO THE ANISOTROPIC PIECE?] the
dominating term comes from what we called the anisotropic bit. it is
anisotropic only in lowest order in $\gd$, but becomes more
isotropic at larger shear rates; (nevertheless, it should alwas
vanish at $q_x=0$, which to me indicates that it's a complicated
function.}

The connection between the dynamic yield contribution in the
stationary structure factor $S_{\qb}(\gd)$ and the actual yield
stress becomes clearer from recognizing a mathematical relation
between \gl{d9} and \gl{d13}, whereby the \mike{shear} stress can be obtained
from an integral over the distorted microstructure: \beq{g9}
\sigma(\gd) = \frac{k_BT \,n}{2} \int\!\!\frac{d^3k}{(2\pi)^3}\;
\frac{k_x\; c'_k \; k_y}{k}\;\; S_\kb(\gd) \; , \eeq where
$c'_k\equiv\partial c_k/\partial k$ enters. The direct correlation
function $c(r)$ plays here the role of the negative of an effective
potential in units of thermal energy. Because of the angular
dependence in \gl{g9}, only the anisotropic contribution $\delta
S^{\rm aniso}_{\bf q}$ enters in the calculation of shear stress,
giving a direct connection between the yield stress contribution to
the distorted structure factor $\delta S^{\rm aniso, + }_{\bf
q}(\varepsilon\ge0)$ and the dynamic yield stress itself. (The
latter is defined as
$\sigma^+(\varepsilon\ge0)=\sigma(\gd\to0,\varepsilon\ge0)$.)

The relation in \gl{g9} was previously found by \citet{Fredrickson}
in a Gaussian field theory for block \rev{copolymers}. We interpret
this similarity as an indication that the ITT approach based on MCT
implicitly assumes a Gaussian distribution for the time-dependent
density fluctuations. This is connected with the breaking of higher
order density moments into products of time-dependent density
correlation functions\rev{, which} lies at the heart of our approach. In
this context it is interesting to note that \citet{Chandler} found
equilibrium density fluctuations in a dense fluid simulation to
follow a Gaussian distribution rather closely even for the highly
nonlinear interaction given by the hard core excluded volume
constraint. It would be very interesting to see this tested under
flow also.

Let us turn finally to the ITT-MCT predictions for the steady shear
stress, which defines the flow curve $\sigma(\gd)$; the shape of
this curve depends solely on $S_q$ (with scale factors in time and
stress deriving from $D_0$, $k_BT$, and the particle diameter $d$).
Numerous such flow curves have by now been obtained, for a variety
of systems with differing $S_q$, either within the ISHSM
\citep{Fuchs2002c,Fuchs2003,Bra:08,Zau:08,Miy:04,Miy:06} or via
fully schematic models like the F$_{12}^{(\gd)}$-model
\citep{Fuchs2003,Cra:06,Cra:08,Var:06b,Haj:08}. Note that exactly
the same ISHSM as used here (Appendix C) forms the basis of Refs.~\citep{Bra:08,Zau:08}; this differs in detail from the earlier
version, as used in
Refs.~\citep{Fuchs2002c,Fuchs2003,Bra:07,Miy:06}. Such differences
are however small, and basically irrelevant compared to the
uncertainty introduced by the unexplained strain scaling parameter
$\gamma_c$ that is generally needed to bring the ISHSM into register
with experimental data. We refer the reader to
Ref.~\citep{Fuchs2003} for a detailed discussion of flow curves from
the original ISHSM and the F$_{12}^{(\gd)}$-model, and to
Refs.~\citep{Bra:08,Zau:08} for additional explicit results.
\disc{IVE NOT DONE ANYTHING BEYOND THIS POINT IN THE PAPER YET,
PENDING BAXTER ETC.}

\section{Conclusions and outlook}
\label{conc}

In \mike{our ITT approach to sheared dense suspensions,} properties
of the stationary state are approximated by following the transient
structural rearrangements encoded in the transient density
correlator $\Phi_{\qb}(t) = \langle \varrho_{\qb}^* \, e^{\smopb
t}\, \varrho_{\qb(t)} \rangle/NS_q = \langle \varrho_{\qb}^* \;
\varrho_{\qb(t)}(t) \rangle/NS_q $. Closed non-Markovian equations
of motion for $\Phi_{\qb}(t)$, obtained after mode coupling
approximations \mike{(}and determined by the equilibrium structure
factor, the density and a short time diffusion coefficient\mike{)}
need to be solved, while stationary averages under shear are
obtained from time- and wavevector integrals containing
$\Phi_{\qb}(t)$. Shear advection of density fluctuations accelerates
the loss of memory.\mikecom{I deleted `whose strength is
approximated to be the equilibrium one' since I don't know what that
means.} \mike{However, this effect} competes with increased
nonlinear coupling of wavevector modes which models the cage-effect
in dense fluids, and which leads to increased memory effects and
relaxation times.

ITT proposes a scenario for the transition between a shear-thinning
viscoelastic  fluid and a yielding/\mike{shear-melted} glass, which
captures many features observed in dense colloidal dispersions. The
physics of the glass transition has thus been \mike{brought to bear on addressing} the
nonlinear rheology of dense dispersions. ITT generalizes the
concept of an ideal glass transition in (quiescent) MCT to driven
(sheared) systems. It gives stationary averages, correlation and
response functions (susceptibilities) \citep{Kru:08}. The essence of the transition
between the two nonequilibrium states \mike{(the elastically distorted glass on the one hand, and the shear-melted state on the other)} can be captured in schematic
models.

Many open questions still remain in the development and application
of \mike{ITT-MCT}. In the formulation of the theory, the neglect of
the stress-induced couplings $\Sigma(t)$ and $\bar \Sigma(t)$ in
\gl{f105} remains unsatisfactory.  We expect it is one cause for the
\mike{theory underestimating the effects of shear, necessitating the
introduction of the strain rescaling parameter $\gamma_c$ to bring
theory and experiment into register. Another contributor to this
mismatch is presumably the isotropic approximation underlying the
ISHSM; and in Fig.\ref{Fig1} a third comes from the fitting of
incoherent stationary correlators to coherent transient ones.
Together these factors require a disturbingly small value of
$\gamma_c$ to be used, and this aspect of the MCT/ITT approach is
not yet fully under control.} The neglect of the stress couplings
leads to another aspect of the theory which should be scrutinized
further. All yield \mike{stress values (the finite stresses
remaining in the limit $\gd\to0$ in the yielding glassy states)} are
determined by the equilibrium structure alone. \mike{Hence they
cannot be affected by e.g.~hydrodynamic modifications to the bare
diffusion constant}. The stress-induced couplings $\Sigma(t)$ and
$\bar \Sigma(t)$ could change this, as seems required to explain
recent simulations of \mike{fluids comprising overdamped Newtonian
particles} under shear \citep{Hor:09}. Differences between the
stationary and the transient density correlators observed in the
same simulation study \citep{Zau:08}, also are not contained in the
present description. Recently, however, an improved approximation
within ITT has been suggested, which comes closer to the measured
data \citep{Kru:08}\mike{, and should in future be} used to
reconsider the stationary correlators from confocal microscopy and
simulation discussed in Section \ref{yield}.

Let us finish in pointing to the recent extension of the
\mike{ITT-MCT} approach to arbitrary incompressible and homogeneous
flows which has led to a general constitutive equation for colloidal
dispersions close to a glass transition \citep{Bra:08}. Consequences
of this theory are being worked out.

\section{Acknowledgements}
  We  acknowledge financial support \mike{of} the
Deutsche Forschungsgemeinschaft in SFB TR6, SFB 513, and IRTG 667, \mike{and of the EPSRC in EP/030173}.
MEC holds a Royal Society Research Professorship. \\

\add{\mike{Note added:} During completion of this manuscript, we
became aware of the work by \citet{Chong08}, who worked out an
ITT-MCT approach for simple fluids under shear as studied in
simulations using the SLLOD equations. A detailed comparison of
\mike{the two} approaches \mike{may prove} very rewarding.}

\appendix
\section{Appendix A}

Appendix A contains some technical material used in the text.

The SO $\smor{t}$ introduced in \gl{e10} is not perpendicular to
density fluctuations and can be analysed using
 \beq{e105'} e^{\dod t} =  e^{\dodb
t} \, \left( 1 + \gd \, \int_0^tdt'\;  e^{-\dodb t'}\, \sigma_{xy}
\,  e^{\dod t'} \right) = e^{\dodb t}\left( 1 + \Sigma(t) \right) \;
. \eeq Here, the difference $\dod-\dodb=\gd\, \sigma_{xy}$ between
the two advection operators becomes important, which is the shear
element of the microscopic stress tensor. Its time integral is
denoted as $\Sigma(t)$ in \gl{e105}. The SO $\smor{t}$, thus can be
decomposed as given in \gl{e108}.

 The
time ordered exponential $u(t,t')$ introduced  in \gl{e11}, \rev{which}
satisfies $\partial_t u(t,t') = u(t,t') a(t)$ with $u(t,t)=1$, and
where $a(t)$ is an operator that does not commute with itself for
different times, is given by (for $t>t'$): \beq{ap0}u(t,t')=
e_-^{\int_{t'}^t ds\; a(s)} = 1 + \sum_{n=1} \int_{t'}^t ds_1\;
\int_{t'}^{s_1} ds_2\;\dots\int_{t'}^{s_{n-1}} ds_{n}\; a(s_n)\,
a(s_{n-1}) \dots a(s_1)\; . \eeq It obeys:
$u(t,t')=u(t'',t')\,u(t,t'')$ for any $t>t''>t'$,
$\partial_{t'}u(t,t')= -a(t') u(t,t')$, and
$u(t,t')^{-1}=v(t,t')=e_+^{-\int_{t'}^t ds\; a(s)}$, where $e_+$ is
built with the reverse order of operators compared to \gl{ap0}, and
solves $\partial_t v(t,t') = -a(t)\, v(t,t')$ with $v(t,t)=1$; see
e.g. \citep{Kawasaki1973b,vankampen}.

The irreducible dynamics appearing in the final formally exact
equation of motion \gl{e21}, can be rewritten using \gl{e165}.  \mike{Standard} manipulations starting from $\partial_t U_i(t,t') =
U_i(t,t')\; \smoi{t}$ lead to
$$U_i(t,t') = U_i^Q(t,t') + \int_{0}^tdt'\; U_i(t',0)\; \smos{t'}\,
\tilde Q(t')\; U_i^Q(t,t')$$ where the newly defined time evolution
operator
$$U_i^Q(t,t') = \exp_-{\{\int_{t'}^tds\; \smoq{s}\; \tilde Q(s)\}}=
 \exp_-{\{\int_{t'}^tds\;
e^{\dodb s}\; Q(s) \; \smoe\; e^{-\dod s}
 \; \tilde Q(s)\}}$$
satisfies $P\; U_i^Q(t,t')= P = U_i^Q(t,t')\; P$. This leads to
\gls{f107}{f2} under the approximation of \gl{f105}.

\section{Appendix B}

The ITT approach introduced by
\citet{Fuchs2002c,Fuchs2005a,Cates2004a}, consisted of equations of
motion slightly different from the ones derived here. Nevertheless,
it led to the identical nonlinear stability equation \gl{g3}  and
thus to the identical universal scenario of a shear thinning fluid
and a yielding glass, which is also the basis \rev{and content} of the
schematic models by \citet{Fuchs2002c,Fuchs2003}. The ISHSM
presented by \citet{Fuchs2002c,Fuchs2003} differed from the one
summarized in Appendix C in quantitative details, \mike{but by an
amount} swamped by the unknown parameter $\gamma_c$ entering because
of isotropic averaging. Thus detailed quantitative comparisons
\mike{between model variants are not warranted. Nonetheless, in this
Appendix}, the differences between the microscopic ITT formulations
of the different papers will be listed to establish an unambiguous
formulation.

A first difference concerns the definition of the transient
correlator, which in version (I) of ITT 
\citep{Fuchs2002c,Fuchs2005a,Cates2004a} carried the wavevector
advection on the left hand side of the average
$$\Phi_\qb^{\rm (I)}(t) = \frac{1}{NS_q}\; \langle \rho^*_{\qb(-t)}\; e^{\smopb t}\; \rho_{\qb} \rangle
 = \frac{S_{q(-t)}}{S_q} \; \Phi_{\qb(-t)}(t) \; ,$$
\rev{where \gl{d1} was used.} This definition, and the present one
in \gl{d1} obey translational invariance and at first sight seem
equivalent. \rev{ Obviously, the sign of $\gd$ is arbitrary, so that
the difference between forward and reverse advection appears
irrelevant.} Yet, \mike{the version (I) correlator} obeys for short
times
$$\Phi_\qb^{\rm (I)}(t\to0) \to  \left( 1 + \frac{1}{S_q} S'_q \frac{q_x q_y}{q} \; \gd t\right)\; \left(1-\frac{q^2}{S_q}\; t\right) + {\cal O}(t^2)= 1 - \frac{t}{S_q}\;\left(q^2-\gd\, S'_q\frac{q_x q_y}{q}\right) + {\cal O}(t^2)\; , $$
which \mike{shows} that the initial decay rate in ITT-(I) is
affected by the stretching of equilibrium density fluctuations
contained in $S_{q(t)}$. \rev{ This subtle difference is the reason
for using the present definition of the advected wavevector in
\gl{b15}.} This complication could be got rid off in ITT-(I) by
performing the limit of $\gd\to0$, accompanied by $t\to\infty$
keeping the strain $t\gd=$ const.. \mike{However,} the present
formulation, where $\gd$ always appears in the fashion of an
accumulated strain $\gd t$, appears simpler. Moreover, it has not
been worked out yet, how the second difference in the ITT equations
of motion, to be discussed next, can be eliminated \mike{from
version (I)}. Therefore, we suggest to use definition \gl{d1} for
the transient correlator from now on.

An important difference to ITT-(I) concerns our present handling of
the linear coupling of density fluctuations to themselves. It leads
to \rev{an} `initial decay rate' $\Gamma_{\qb}(t)$ that becomes
time-dependent under shear. In ITT-(I), $\Gamma^{(I)}_{\qb}(t)$
could exhibit different signs depending on the direction of $\qb$.
In numerical solutions of ITT-(I), the appearance of
$1/\Gamma^{(I)}_{\qb}(t)$ as prefactor in the memory kernel
destabilized the numerical algorithm at long times, restricting the
density and shear rate window where ITT-(I) could be applied
\citep{Hen:07}. The stability of the equations of motion derived by
\citet{Fuchs2002c,Fuchs2005a,Cates2004a} could not be assured.
\rev{While stable equations might be possible with the old
definition of the transient correlator, the new definition in
\gl{d1} easily led to the convenient time evolution operator in
\gl{e6}. Again, it appears not to be the only choice giving stable
relaxations, but serves well at least as long as the accumulated
elastic energies $\Sigma(t)$ are neglected.} Introduction of the
advection operator $\dod$, working symmetrically with collective
densities, and the special choice of projection in \gl{e9}
\mike{overcomes} the problem of negative $\Gamma_{\qb}(t)$, and
\mike{leads} to the vertices in the memory function of \gl{f8},
which assure a positive instantaneous friction. Therefore, we
suggest to use the approximation \gl{f8} for the memory kernel from
now on. Quantitatively, but not qualitatively, the results on
step-strain by \citet{Bra:07} need to be \mike{recalculated},
because there the details of the ITT-(I) vertex mattered. All other
results presented in Refs.~\citep{Fuchs2002c,Fuchs2003,Fuchs2005a,Cates2004a,Bra:07} remain unaffected, as  do all results in Refs.~\citep{Fuchs2002c,Fuchs2003,Cra:06,Cra:08,Haj:08,Var:06b} based on
schematic models. \mike{However}, as already mentioned, the result
\mike{given} for the isotropic term in the distorted structure
factor by \citet{Henrich2007} was inaccurate \mike{as it assumes
consistency at large $q$ which ITT-MCT in fact does not achieve.}
\com{Add a bit about why there is no $\Delta$ in old formulation.}
\mikecom{You wanted to add a bit about why there is no $\Delta$ in
old formulation. This would be good if it can be explained clearly
and briefly. I'm not sure how to do that!}

\section{Appendix C}

For completeness, the numerically solved ISHSM is summarized. It
consists of \gl{g1} with the assumption of isotropic correlators,
$\Phi_q(t)$. The initial decay rate is approximated without shear:
$\Gamma_q=q^2 D_s / S_q$, where $D_s$ is the density-dependent short
time diffusion coefficient. The memory function also is taken as
isotropic and modeled close to the unsheared situation \beq{iso2}
 m_q(t) \approx
\frac{1}{2N} \sum_{\kb}  V^{(\gd)}_{\qb,\kb}(t)\;
 \Phi_k(t) \, \Phi_{|\qb-\kb|}(t) \; ,
\eeq with \beq{iso3} V^{(\gd)}_{\qb,\kb}(t) =
\frac{n^2S_qS_kS_p}{q^4} \left[ \qb\cdot\kb\; c_{\bar k(t)} +
\qb\cdot\pb\; c_{\bar p(t)}  \right] \left[ \qb\cdot\kb\; c_{k} +
\qb\cdot\pb\; c_{p} \right] \eeq where $\pb=\qb\!-\!\kb$, and the
length of the advected wavevectors is approximated by $\bar
k(t)=k(1+(t\gd/\gamma_c)^2 )^{1/2}$ and  $\bar
p(t)=p(1+(t\gd/\gamma_c)^2 )^{1/2}$ . Note, that the memory
function thus only depends on one time. We use the \mike{phenomenological adjustment} factor $\gamma_c$
in order to correct for the underestimate of the effect of shearing in the ISHSM.
\begin{figure}[t]
\centering
\includegraphics[ width=0.6\columnwidth]{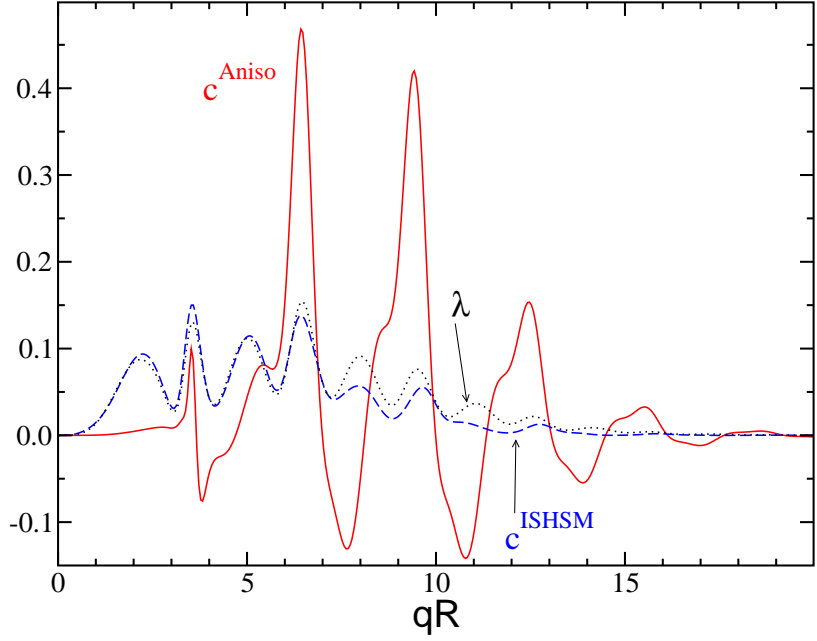}
\caption{Integrands for the parameter $c^{(\gd)}$ in the stability
equation (\ref{g3}); the one labeled $c^{\rm Aniso}$ results for the
memory function in \gl{f8}, the one labeled $c^{\rm ISHSM}$ for the
ISHSM approximation \gl{iso2} setting $\gamma_c=0.84$ to recover
$c^{(\gd)}=0.64$ in both \mike{cases}. For comparison the
integrand leading to the exponent parameter $\lambda$ is included as
dotted line. \label{Fig4}}
\end{figure}

The expression  for the potential part of the transverse stress may be simplified to
\beq{iso4} \sigma\approx\frac{k_BT\, \gd}{60\pi^2}\;
\int_0^\infty\!\! dt\; \int\!\! dk\;
  \frac{k^5\, c'_k\; S'_{\bar k(t)}}{\bar k(t)}\; \Phi_{\bar k(t)}^2(t) \; ,
\eeq while the (anisotropic) structure factor is approximated as
\beqa{iso5}
\delta S^{\rm aniso}_\qb(\gd) &\approx&  \gd q_x q_y \, \int_0^\infty dt\; \frac{S'_{\bar q(t)}}{\bar q(t)}\; \Phi_{\bar q(t)}^2(t) \quad \mbox{for }\; \gd\tau \ll 1 \; ,\nonumber\\
\delta S^{\rm aniso}_\qb(\gd) &\approx&  q^2 \frac{\gd^2
}{\gamma_c^2} \, \int_0^\infty dt\; \frac{t S'_{\bar q(t)}}{\bar
q(t)}\; \Phi_{\bar q(t)}^2(t)\quad \mbox{for }\; \gd \tau \gg 1
 \; . \eeqa In the last two equations, \gls{iso4}{iso5},
the parameter $\gamma_c$ is taken to be $\gamma_c=\sqrt{3}$, as
would follow from isotropic averaging of $\kb(t)$. Note that in the
glass, the distorted structure factor becomes isotropic in
ISHSM\add{, because for $\gd\tau \gg1$, the isotropization
approximation should be performed on \gl{g7}, which leads to the
second line in \gl{iso5}}.

For the numerical solution of the ISHSM the algorithm by
\citet{Fuc:91} was used for hard spheres with radius $R$. Their
structure factor $S_k$ is taken from the Percus-Yevick approximation
\citep{russel} and depends only on the packing fraction $\phi$.  The
wavevector integrals were discretized according to \citet{Fra:97}
with $M=100$ wavevectors from $k_{\rm min}=0.1/R$ up to $k_{\rm
max}=19.9/R$ with separation $\Delta k = 0.2/R $. Time was
discretized with initial step-width $dt=2\rev{\times}10^{-7} R^2/D_0$, which
was doubled each time after 400 steps. We find that the model's
glass transition lies at $\phi_c=0.51591$. Thus
$\epsilon=(\phi-\phi_c)/\phi_c$ (where $\tilde{\epsilon} \simeq
1.54\, \epsilon$), and $\gd$ are the only two control parameters
determining the rheology. The exponent parameter becomes
$\lambda=0.735$ and $c^{(\gd)}\approx 0.45 / \gamma_c^2$ ; note that
these values still change somewhat if the discretization is made
finer.

The ISHSM value for $c^{(\gd)}$ gives a rough estimate of the error
in ISHSM caused by using  isotropically averaged vertices. Without
this approximation, $c^{(\gd)}\approx 0.64$ holds for the present
discretization in an anisotropic calculation. This leads to the
conclusion that $\gamma_c\approx0.84$ should approximately correct
for the error when isotropically averaging. Yet, as discussed in
Section \ref{yield} appreciably smaller values $\gamma_c\le0.1$ are
required to fit the measured data.
 More information on the spatial
distribution of the error in the ISHSM can be gained from
considering the integrands in the expressions leading to
$c^{(\gd)}=\int dq\, i^{c}_q$ derived by \citet{Fuchs2003}. They are
shown in Fig.~\ref{Fig4} for the ISHSM using the value
$\gamma_c=0.84$, and for the anisotropic calculation. The ISHSM not
only errs in magnitude, \mike{(e.g.~ in} the prefactor of the shear
induced relaxation time\mike{)} but also qualitatively misses the
negative contributions. The curve with anisotropic shear-advection
taken properly into account indicates that the particles rearrange
locally during the initial part of the yield process, as
contributions from the range $2\le qR\le 20$ dominate.

\end{document}